\documentclass[fleqn,10pt, twocolumn]{wlscirep}
\usepackage{multirow}

\usepackage{subcaption}
\usepackage{wrapfig}
\usepackage{mathrsfs}
\usepackage{hyperref}
\usepackage{nameref}
\usepackage[normalem]{ulem}

\usepackage{multicol}

\newcommand{\beginsupplement}{%
        \setcounter{table}{0}
        \renewcommand{\thetable}{\arabic{table}}%
        \setcounter{figure}{0}
        \renewcommand{\thefigure}{\arabic{figure}}%
        \renewcommand{\figurename}{Supplementary Figure} 
        \renewcommand{\tablename}{Supplementary Table} 
     }

\title{
Robust recognition and exploratory analysis of crystal structures via Bayesian deep learning
}
 
\author[1,*]{Andreas Leitherer}
\author[1]{Angelo Ziletti}
\author[1]{Luca M. Ghiringhelli}
\affil[1]{Fritz-Haber-Institut der Max-Planck-Gesellschaft, 14195 Berlin-Dahlem, Germany}

\affil[*]{leitherer@fhi-berlin.mpg.de}

 \begin{abstract}
Due to their ability to recognize complex patterns, neural networks can drive a paradigm shift in the analysis of materials science data. 
Here, we introduce ARISE, a crystal-structure identification 
method based on Bayesian deep learning. 
As a major step forward, 
ARISE is robust to structural noise and can treat more than 100 crystal structures, a number that can be extended on demand. 
While being trained on ideal structures only, ARISE correctly characterizes strongly perturbed single- and polycrystalline systems, from both synthetic and experimental resources.
The probabilistic nature of the Bayesian-deep-learning model allows to obtain principled uncertainty estimates, which are found to be correlated with crystalline order of metallic nanoparticles in electron tomography experiments. 
Applying unsupervised learning to the internal neural-network 
representations reveals grain boundaries and (unapparent) structural regions sharing easily interpretable geometrical properties. 
This work enables the hitherto hindered analysis of noisy atomic structural data from computations or experiments.
 \end{abstract}
\begin{document}

\flushbottom
\maketitle

\thispagestyle{empty}

\section*{Introduction}

Identifying the crystal structure of a given material is important for understanding and predicting its physical properties.
For instance, the hardness of industrial steel is strongly influenced by the atomic composition at grain boundaries, which has been studied in numerous
theoretical and experimental investigations\cite{herbig2014atomic, meiners2020observations}. 
Beyond bulk materials, two- (2D) and one-dimensional (1D) systems have far-reaching technological applications, such as solar energy storage, DNA sequencing, cancer therapy, or even space exploration\cite{ferrari2015science, de2013carbon}.  
To characterize the crystal structure of a given material, one may assign a symmetry label, e.g., the space group.
More generally, one may want to find the most similar structure within a list of given known systems. These so-called structural classes are identified 
by stoichiometry, space group, number of atoms in the unit cell, and location of the atoms in the unit cell (the Wyckoff positions). 

  Methods for automatic crystal-structure recognition are required to analyze the continuously growing
 amount of geometrical information on crystal structures, from both experimental and computational studies.
Millions of crystal structures alongside calculated properties are available in large computational databases such as
  the NOvel MAterials Discovery (NOMAD) Laboratory \cite{draxl2019nomad},  
 AFLOW\cite{mehl2017aflow},  
 the Open Quantum Materials Database (OQMD)\cite{saal2013materials}, Materials Project\cite{jain2011high}, 
 or repositories specialized in 2D materials\cite{haastrup2018computational,mounet2018two}. 
In scanning transmission electron microscopy (STEM)\cite{pennycook2011scanning}, atomic positions can be reconstructed from atomic-resolution images 
for specific systems, e.g., graphene \cite{ziatdinov2017deep}. 
Three-dimensional atomic positions 
are provided by atom probe tomography (APT) \cite{gault2012atom} and atomic electron tomography (AET) experiments\cite{zhou2020atomic}. 
Still, substantial levels of noise due to experimental limitations and reconstruction errors are present in atomic positions, e.g., distortions beyond a level that can 
be explained by a physical effect or, 
in case of APT, large amount of missing atoms (at least $20\%$, due to the limited detector efficiency\cite{gault2016brief}). 
Crystal-structure recognition schemes should be able to classify a large number of structural classes 
(also beyond bulk materials) while at the same time being robust with respect to theoretical or experimental sources of inaccuracy and 
 physically driven deviations from ideal crystal symmetry (e.g., vacancies or thermal vibrations). 
Given the large amount of data, the classification should 
be fully automatic and independent of the manual selection of tolerance parameters (which quantify the deviation from an ideal reference structure). 
Current methods are based either on space-group symmetry or local structure. 
For space-group-based approaches (notable examples being Spglib\cite{togo2018texttt} and AFLOW-SYM\cite{hicks2018aflow}), the allowed symmetry operations are calculated directly from the atomic 
positions to infer a space group label. 
For local-structure-based approaches, the local atomic neighborhood of each individual atom is classified into a predefined list of reference structures.  
Examples of these methods are common neighbor analysis (CNA)\cite{honeycutt1987molecular},
adaptive common neighbor analysis (a-CNA)\cite{stukowski2012structure}, bond angle analysis (BAA)\cite{ackland2006applications}, and  polyhedral template matching (PTM)\cite{larsen2016robust}. 
Space-group approaches can treat all space groups but are sensitive to noise, while local-structure methods can be quite robust but only treat a handful of structural classes. 
Moreover, none of the available structure recognition schemes can recognize 
complex
nanostructures, e.g., nanotubes. 

To improve on the current state of the art, we build on recent advances in deep learning, which is a subfield of machine learning that yields ground-breaking results in many settings, e.g., image and 
speech recognition\cite{Goodfellow-et-al-2016}. 
Previous work using machine learning and neural networks (NNs) for crystal-structure recognition\cite{geiger2013neural,reinhart2017machine,dietz2017machine,ziletti2018insightful} 
did not go beyond a handful of structural classes while showing robustness at the same time. 

Here, we propose a robust, threshold-independent crystal-structure 
recognition framework (ARtificial-Intelligence-based Structure Evaluation, short ARISE) to classify a diverse set of 108 structural classes, comprising bulk, 2D, and 1D materials. 
Bayesian NNs\cite{gal2016dropout,gal2016uncertainty} are used,
i.e., a recently developed family of NNs that yields not only a classification but also uncertainty estimates. These estimates are principled in the sense that they approximate those of a well-known probabilistic model (the Gaussian process). 
This allows to quantify prediction uncertainty, but also the degree of crystalline order in a material.
ARISE performance is compared with the current state of the art, and then applied to various computational and experimental atomic structures. 
Crystal characterization and identification of hidden patterns is performed using supervised learning (ARISE) as well as the unsupervised analysis (via clustering and dimensionality reduction) of the internal representations 
of ARISE.

\section*{Results}

 \begin{figure*}
 \centering
 \includegraphics[width=\textwidth]{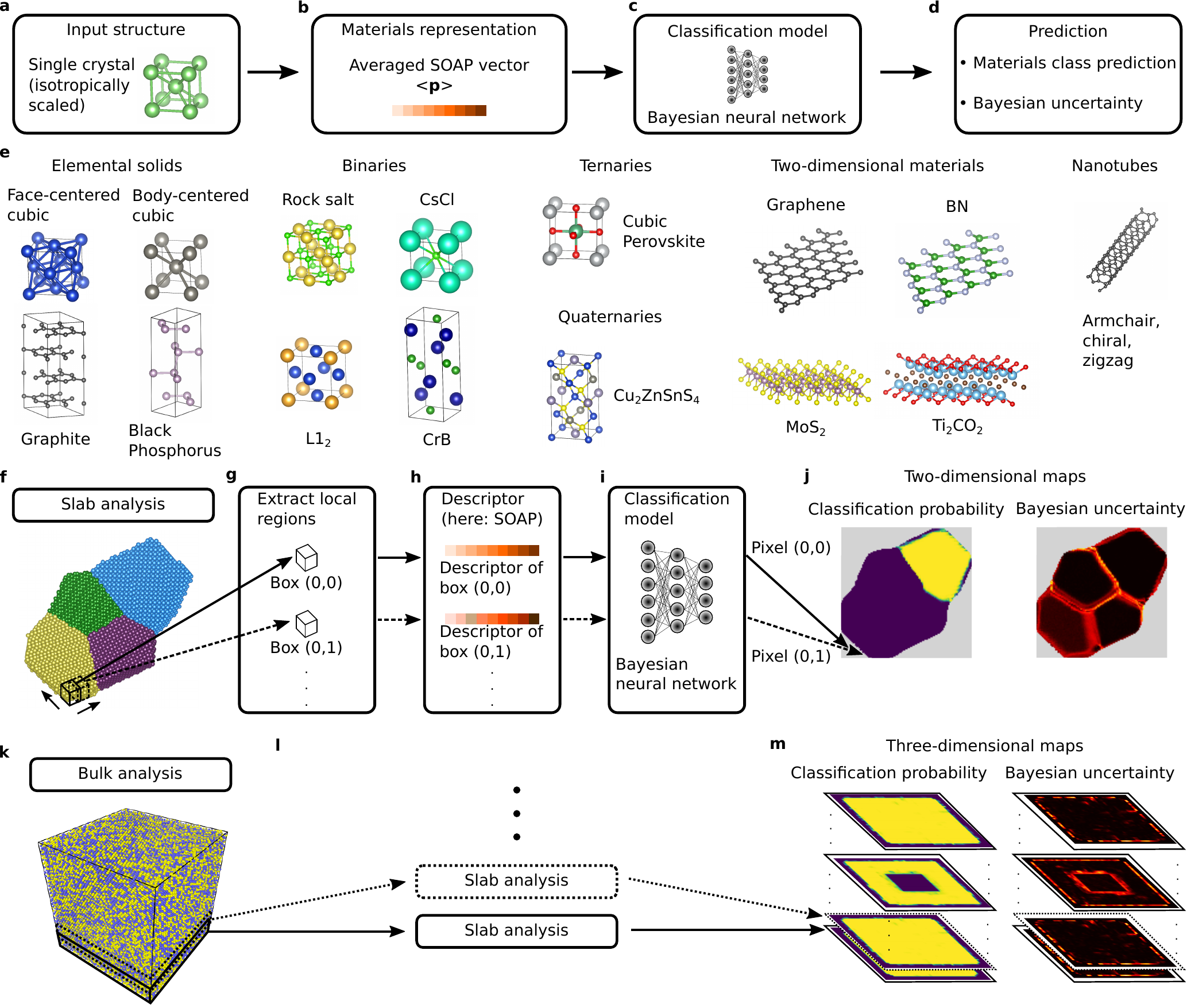}
 \caption{
 \textbf{Schematic overview of single- and polycrystal characterization framework.} \textbf{a-d} Prediction pipeline of the 
 single-crystal classification model ARISE (ARtificial-Intelligence-based Structure Evaluation). 
 In this work we employ the smooth-overlap-of-atomic-positions (SOAP) descriptor. 
 \textbf{e} Examples of crystallographic prototypes included in the training set. \textbf{f-m} Polycrystal classification framework strided pattern matching (SPM) for slab-like (\textbf{f-j}) and bulk systems (\textbf{k-m}).
 }
 \label{fig:single_and_polyc_class_steps}
 \end{figure*}

\subsection*{The input representation}
To apply machine learning to condensed-matter and materials science problems, the input coordinates, chemical species, and the lattice periodicity of a 
given atomic structure are mapped onto a suitable so-called descriptor. Here, the descriptor is a vector that is invariant under rigid translations and rotations of the input 
structure, as well as under permutations of same-species atoms. 
Quality and generalization ability of machine-learning models can be significantly increased, if physical requirements known to be true  
are respected by construction (see Supplementary Methods  
for more details).

Most well-known descriptors in physics and materials science incorporate these
physical invariants: 
symmetry functions\cite{behler2011atom}, the smooth-overlap-of-atomic-positions descriptor (SOAP) \cite{bartok2010gaussian,bartok2013representing}, 
 the many-body tensor representation\cite{huo2017unified}, and the moment tensor potential representation\cite{shapeev2016moment}. 
In this work, SOAP is used as descriptor (cf. Supplementary Methods).   
SOAP has been successfully applied to numerous materials science problems such as interatomic potentials fitting \cite{bartok2015gaussian}, structural similarity quantification\cite{de2016comparing}, 
or prediction of grain boundary characteristics (e.g., energy and mobility)\cite{rosenbrock2017discovering}.
Note that any other suitable descriptor that respects above-mentioned physical requirements can be used as input for our procedure. 
In particular, the ai4materials code library is provided into which alternative descriptors can be readily integrated.

\subsection*{The Bayesian deep learning model and the training dataset}

Once the crystal structures are converted into vectorial descriptors by means of the SOAP mapping, a NN model is used to arrive at a classification decision (cf. Fig. \ref{fig:single_and_polyc_class_steps}c). 
 NNs are nonlinear machine-learning models: they transform the input in a hierarchical fashion by subsequently applying affine and non-linear transformations in a predefined series of layers. 
The NN learns these optimal transformations that deform the descriptor space so that a robust classification is achieved. 
In this way, the model is able to learn complex representations which are becoming more abstract from layer to layer\cite{ziletti2018insightful}. 
This ability to learn representations\cite{bengio2013representation} is one of the key characteristics distinguishing NNs from other machine-learning algorithms.  
Various NN architectures have been developed in recent years\cite{Goodfellow-et-al-2016}; 
in this work, a fully connected NN (multilayer perceptron) is employed. 

 A key component of this work is something rarely addressed in machine learning applied to materials science: quantification of model prediction uncertainty (cf. Fig. \ref{fig:single_and_polyc_class_steps}d). 
 Standard NNs are unable to provide reliable model uncertainty \cite{gal2016dropout}.
In a classification setting, there is widespread use of the probability provided by the last layer as uncertainty estimate.
These probabilities are typically obtained by normalizing the sum of output values using the so-called softmax activation function. 
 The  class with maximal probability corresponds to the final prediction (here of a specific structural class). One may interpret the classification probability as quantification of model confidence.  
However, this strategy is unreliable as 
standard NNs tend to erroneously assign unjustified high confidence to points
for which a low confidence should be returned instead\cite{gal2016dropout}.
The main reason for this behavior is that standard-NN predictions are deterministic, with the softmax output only providing point estimates of the true probability distribution of outputs.    
In Bayesian NNs, this is addressed by placing distributions over model parameters. This results in probabilistic outputs \textendash\ in contrast 
to the point estimates from deterministic NNs\textendash\ from which principled uncertainty estimates can be obtained.
Gal and Ghahramani\cite{gal2016dropout} showed that high-quality uncertainty estimates (alongside predictions) can be calculated 
at low cost using stochastic regularization techniques  
such as dropout\cite{hinton2012improving, srivastava2014dropout} (see Supplementary Methods  
for more details).

After both descriptor and model architecture have been identified, a diverse, comprehensive, 
and materials-science-relevant training set is constructed. The first  \textendash\ and 
most important  \textendash\ step is to define the structural classes which are going to be included in the model:  
an overview of the structural classes considered in this work is shown in 
Fig. \ref{fig:single_and_polyc_class_steps}e. 
This comprehensive collection of structures includes bulk materials of elemental, binary, ternary, and quaternary composition, as well as 2D materials and carbon nanotubes of chiral, armchair, and zigzag type. 
In practice, given any database, we extract prototypes, i.e., representative structures that are selected according to some predefined rules.
Selection criteria are, for instance, fulfillment of geometrical constraints (number of atoms in the unit cell, number of chemical species) or if the structures are observed in experiment.
For the elemental bulk materials, we extract from AFLOW all experimentally observed structures with up to four atoms in the primitive cell. 
This yields 27 elemental solids encompassing all Bravais lattices, with the exception of monoclinic and triclinic structures because of their low symmetry.
Note that this selection includes not only the most common structures such as 
face-centered-cubic (fcc), body-centered-cubic (bcc), hexagonal-close-packed (hcp), and diamond (which cover more than $80\%$ of the elemental solids found in nature\cite{ashcroft2011solid}), 
but also double-hexagonal close-packed, graphite (hexagonal, rhombohedral, buckled), and orthorhombic systems such as black phosphorus. 
This goes already beyond previous work using NNs for crystal structure recognition\cite{ziletti2018insightful}, where a smaller set of elemental solids is considered.
For binaries, we select the ten most common binary compounds according to Pettifor\cite{pettifor1995bonding}, plus the $\text{L1}_{2}$ structure 
because of its technological relevance \textendash\ for instance, it being the crystal structure of common precipitates in Ni-based superalloys\cite{reed2008superalloys}. 
This selection also include non-centrosymmetric structure, i.e. structures without inversion symmetry, such as wurtzite.
To challenge the classification method with an increasing number of chemical species, a small set of ternary and quaternary materials is included as a proof-of-concept.
Specifically, six ternary perovskites\cite{castelli2015calculated} (organometal halide cubic and layered perovskites) and six quaternary chalcogenides of $\text{A}_2\text{BCX}_4$
type\cite{pandey2018promising} are included due to their relevance in solar cells and photo-electrochemical water splitting devices, respectively.
Going beyond bulk materials, we add an exhaustive set of 46 2D materials, comprising not only the well-known elemental structures 
 such as graphene and phosphorene\cite{novoselov20162d} 
but also binary semiconductors and insulators (BN, GaN), transition metal dichalcogenides (MoS$_2$), and one example of metal-organic 
perovskites with six different chemical species.  
Ternary, quaternary, and 2D materials are taken from the computational materials repository (CMR)\cite{landis2012computational}.
To demonstrate the ability of the proposed framework to deal with complex nanostructures, 12 nanotubes of armchair, chiral, and zigzag type are included in the dataset. 
For each prototype, we calculate the SOAP vector with different parameter settings (see Supplementary Methods 
for more details) 
as well as periodic and non-periodic boundary conditions to have a 
comprehensive dataset to train a robust classification model. 
This results in 39$\,$204 (pristine) structures included in the training set.

To optimize the model, the set of pristine structures is split, with $80\%$ being used for training and the remaining $20\%$ for validation. 
For hyperparameter tuning, we employ Bayesian optimization\cite{10.5555/3042817.3042832}, which allows 
to optimize functions whose evaluation is computationally costly, making it particularly attractive for deep-learning models. Here, hyperparameters such as  
learning rate or number of layers are optimized in an automatic, reproducible, and computationally efficient manner to minimize the validation accuracy. 
A list of candidate models is then obtained, from which the optimal model is selected (see Methods section). We term this model ARISE, and report its architecture in Table \ref{table:mlp_all_data}.

\subsection*{Benchmarking}
We now compare ARISE's performance on pristine and defective structures with state-of-the-art crystal-structure recognition methods, 
specifically spglib, CNA, a-CNA, BAA, and PTM (cf. Table \ref{table:accuracy-comparison-single-crystal}). 
As mentioned in the Introduction, none of the benchmarking methods can treat all the materials shown in Fig. \ref{fig:single_and_polyc_class_steps}e; 
thus for fairness, the classification accuracy is only calculated for classes for which the respective methods were designed for, implying that most structures are excluded (see \nameref{section:supp_note_1}  
for more details).

The performance on pristine structures is reported in Table \ref{table:accuracy-comparison-single-crystal}. 
The accuracy in classifying pristine structures is always 100\% as expected, with the only exception being CNA: For this method, the default cutoff only allows to correctly classify fcc and bcc but not hcp structures. 
For defective structures, the situation is drastically different. 
Spglib classification accuracy on displaced structures is low,
and only slightly improved by using loose setting (up to $1\%$ displacement). 
For missing atoms, the accuracy is very low already at the $1\%$ level regardless of the setting used. 
Note, however, that this is actually spglib's desired behavior 
since the aim of this method is not robust classification. 
As indicated in the first column of Table \ref{table:accuracy-comparison-single-crystal}, 
spglib can treat 96 out of the 108 prototypes included in our dataset with the twelve missing prototypes being carbon nanotubes. 
Methods based on local atomic environments (PTM, BAA, CNA, a-CNA) perform very well on displaced structures, 
but they suffer from a substantial accuracy drop for missing-atoms ratios beyond 1\%. Their biggest drawback, however, is that they can treat only a handful of classes: three classes for BAA, CNA, and a-CNA, and twelve classes for PTM.
 ARISE is very robust with respect to both displacements and missing atoms (even concurrently, cf. Supplementary Table \ref{table:suppl_vac_and_displ}), 
while being the only method able to treat all 108 classes included in the dataset,  
including complex systems, such as carbon nanotubes. 
An uncertainty value quantifying model confidence is also returned,
which is particularly important when investigating defective structures or inputs that are far out of the training set. 
We provide a detailed study in \nameref{section:supp_note_3} and Supplementary Fig. \ref{fig:assign_most_sim_proto},  
where 
we challenge ARISE with structures it has not been trained on, i.e., it is forced to fail by construction. We find that ARISE returns 
non-trivial physically meaningful predictions, thus making it particularly attractive, e.g., for screening large and structurally diverse databases. 
Moreover, we analyze predictions and uncertainty of ARISE for continuous structural transformations (cf. \nameref{section:supp_note_2} and Supplementary Fig. \ref{fig:Bain_path_resuts}),  
where we consider the so-called Bain path that includes transitions between fcc, bcc, and tetragonal structures.
 We also want to emphasize that compared to available methods, the classification via ARISE does not require any threshold specifications (e.g., precision parameters as in spglib).

 \begin{table*}[ht]
\centering
\begin{tabular}{ll}
\hline
Layer type & Specifications \\
\hline
\hline
Input Layer & Materials representation \\
+ Dropout  & (SOAP descriptor, size: 316)\\
Fully connected layer & Size: 256 \\
+ Dropout + ReLU  & \\
Fully connected layer & Size: 512\\
+ Dropout + ReLU  & \\
Fully connected layer & Size: 256\\
+ Dropout + ReLU  & \\
Fully connected layer & Size: 108 (= \# classes) \\ 
+ Softmax  & \\
\hline
\end{tabular}
\caption{Architecture of the fully connected Bayesian neural network used in this work. 
Rectified Linear Unit (ReLU) activation functions are used for all hidden layers. 
The dropout ratio is 3.17\% for all layers. 
The total number of parameters is 371,820. While training time was fixed to 300 epochs, hyperopt found a batch size of 64 and  a learning rate of 2.16$\cdot10^{-4}$.}
\label{table:mlp_all_data}
\end{table*}
 
\begin{table*}[]
\begin{tabular}{@{}lrrrrrrrrrrrr@{}}
\hline \hline
& \multicolumn{1}{c}{Pristine} & \multicolumn{1}{l}{} & \multicolumn{5}{c}{Random displacements ($\delta$)}                                                                                     & \multicolumn{1}{l}{} & \multicolumn{4}{c}{Missing atoms ($\eta$)} \\ 
\cmidrule(lr){4-8} \cmidrule(l){10-13} & & & \multicolumn{1}{c}{0.1\%}  & \multicolumn{1}{c}{0.6\%} & \multicolumn{1}{c}{1\%} & \multicolumn{1}{c}{2\%} & \multicolumn{1}{l}{4\%} & \multicolumn{1}{l}{} & \multicolumn{1}{c}{1\%} &  \multicolumn{1}{c}{5\%} & \multicolumn{1}{c}{10\%} & \multicolumn{1}{c}{20\%}\\ 
\cmidrule(r){1-8} \cmidrule(l){9-13} 
Spglib, loose  & 100.00&& 100.00  & 100.00  & 95.26  & 20.00 &  0.00 && 11.23 &  0.00  & 0.00 & 0.00   \\
(96 / 108) & &&   &   &   &  &   &&  &    &  &    \\
 & &&   &   &   &  &   &&  &    &  &    \\
Spglib, tight  & 100.00 &&  0.00 &  0.00 &  0.00 & 0.00 &  0.00 && 11.23 & 0.00   & 0.00  & 0.00  \\
(96 / 108) & &&   &   &   &  &   &&  &    &  &    \\
 & &&   &   &   &  &   &&  &    &  &    \\
PTM   &100.00 && 100.00  & 100.00   & 100.00  & 100.00  &   100.00 && 88.67 & 51.76 &   25.93 & 6.24\\ 
(12 / 108) & &&   &   &   &  &   &&  &    &  &    \\
 & &&   &   &   &  &   &&  &    &  &    \\
CNA  & 66.14 && 62.81  & 62.81   & 54.55  & 32.34  &  31.41 && 55.86  & 32.50    & 15.75 & 3.07   \\
(3 / 108) & &&   &   &   &  &   &&  &    &  &    \\
 & &&   &   &   &  &   &&  &    &  &    \\
a-CNA  & 100.00 && 100.00  & 100.00   & 100.00  & 100.00  &  100.00 && 89.25  & 52.81    & 25.92 & 5.37   \\
(3 / 108) & &&   &   &   &  &   &&  &    &  &    \\
 & &&   &   &   &  &   &&  &    &  &    \\
BAA  & 100.00 && 100.00  & 100.00   & 100.00  & 100.00  &  97.85 && 99.71  & 88.78    & 65.21 & 25.38   \\
(3 / 108) & &&   &   &   &  &   &&  &    &  &    \\
 & &&   &   &   &  &   &&  &    &  &    \\
GNB  &62.63 && 56.50 & 55.94 & 55.56& 54.98  & 52.72  &&  54.51  &52.94 & 52.67 & 52.09 \\
(108 / 108) & &&   &   &   &  &   &&  &    &  &    \\
 & &&   &   &   &  &   &&  &    &  &    \\
BNB  &75.76 && 65.56 & 65.19 & 63.61& 61.58  & 56.58  &&  65.49  &64.00 & 62.43 & 60.48 \\
(108 / 108) & &&   &   &   &  &   &&  &    &  &    \\
 & &&   &   &   &  &   &&  &    &  &    \\
ARISE  &100.00 && 100.00 & 100.00 & 100.00& 99.86  & 99.29  &&  100.00  &100.00 & 100.00 & 99.85  \\
(108 / 108) & &&   &   &   &  &   &&  &    &  &    \\

\hline \hline
\end{tabular}
\caption{\textbf{Accuracy in identifying the parent class of defective crystal structures.} The defective structures are generated by randomly displacing atoms 
according to a uniform distribution on an interval  $\left[ - \delta \cdot d_{\rm NN}, + \delta \cdot d_{\rm NN} \right]$ proportional 
to the nearest neighbor distance $d_{\rm NN}$ (central panel), or removing $\eta$\% of the atoms (right panel). The accuracy values shown are in percentage.  
For benchmarking we use Spglib\cite{togo2018texttt} (with two settings for the precision parameters, so-called loose (position/angle tolerance 0.1\AA/ 5$^\circ$) and tight (position/angle tolerance $10^{-4}$ / 1$^\circ$)), 
polyhedral template matching (PTM)\cite{larsen2016robust}, common neighbor analysis (CNA)\cite{honeycutt1987molecular}, adaptive common neighbor analysis (a-CNA)\cite{stukowski2012structure}, and bond angle analysis (BAA)\cite{ackland2006applications}. 
The number of classes which can be treated out of 
the materials pool in Fig. \ref{fig:single_and_polyc_class_steps}e is shown in parentheses for each method. spglib can assign a space group to all materials except the 12 nanotubes. 
PTM can only classify 7 elemental and 5 binary materials of those considered in this work.
Additional classes are missing for  
CNA, a-CNA, and BAA as they cannot classify simple cubic (sc) and diamond structures.
The approach proposed here can be applied to all classes,
and thus the whole dataset is used (see Supplementary Tables \ref{table:prototype_listing_part_I}-\ref{table:prototype_listing_part_III} for a complete list). 
 Moreover, we compare ARISE to a standard Bayesian approach: Naive Bayes (NB). We consider two different variants of NB: Bernoulli NB (BNB) and Gaussian NB (GNB) \textendash\ see the Methods section 
for more details. ARISE is 
overwhelmingly more accurate 
than both NB methods, for both pristine and defective structures.
} 
\label{table:accuracy-comparison-single-crystal}
\end{table*}

\subsection*{Polycrystal classification}
Up to this point, we have discussed only the analysis of single-crystal (mono-crystalline) structures, using ARISE. 
To enable the local characterization of polycrystalline systems,
we introduce strided pattern matching (SPM). 
For slab-like systems (cf. Fig. \ref{fig:single_and_polyc_class_steps} f), a box of predefined size is scanned in-plane across the whole crystal with a given stride; 
at each step, the atomic structure contained in the box is represented using a suitable descriptor (cf. Fig. \ref{fig:single_and_polyc_class_steps} g-h),
and classified 
(Fig. \ref{fig:single_and_polyc_class_steps}i), yielding a collection of classification probabilities (here: 108) 
with associated uncertainties. 
These are arranged in 2D maps (Fig. \ref{fig:single_and_polyc_class_steps}j). 
The classification probability maps indicate how much a given polycrystalline structure locally resembles a specific crystallographic prototype.
The uncertainty maps quantify the statistics of the output probability distribution (cf. Supplementary Methods).  
Increased uncertainty indicates that the corresponding local segment(s) deviates from the prototypes known to the model. 
Thus, these regions are likely to contain defects such as grain boundaries, or more generally atomic arrangements different from the ones included in training. 
For bulk systems (Fig. \ref{fig:single_and_polyc_class_steps}k), the slab analysis depicted in Fig. \ref{fig:single_and_polyc_class_steps}f-j is repeated for multiple slices (Fig. \ref{fig:single_and_polyc_class_steps}l),
resulting in 3D classification probability and uncertainty
maps (Fig. \ref{fig:single_and_polyc_class_steps}m).

SPM extends common approaches such as labeling individual atoms with symmetry labels\cite{stukowski2012structure}, as the striding allows to discover structural transitions within polycrystals in a smooth way. 
SPM can be applied to any kind of 
 data providing atomic positions and chemical species.   
Results obtained via SPM are influenced  by the quality of the classification model as well as box size and stride (see Methods section for more details).

\subsection*{Synthetic polycrystals}

\begin{figure*}
  \includegraphics[width=\textwidth]{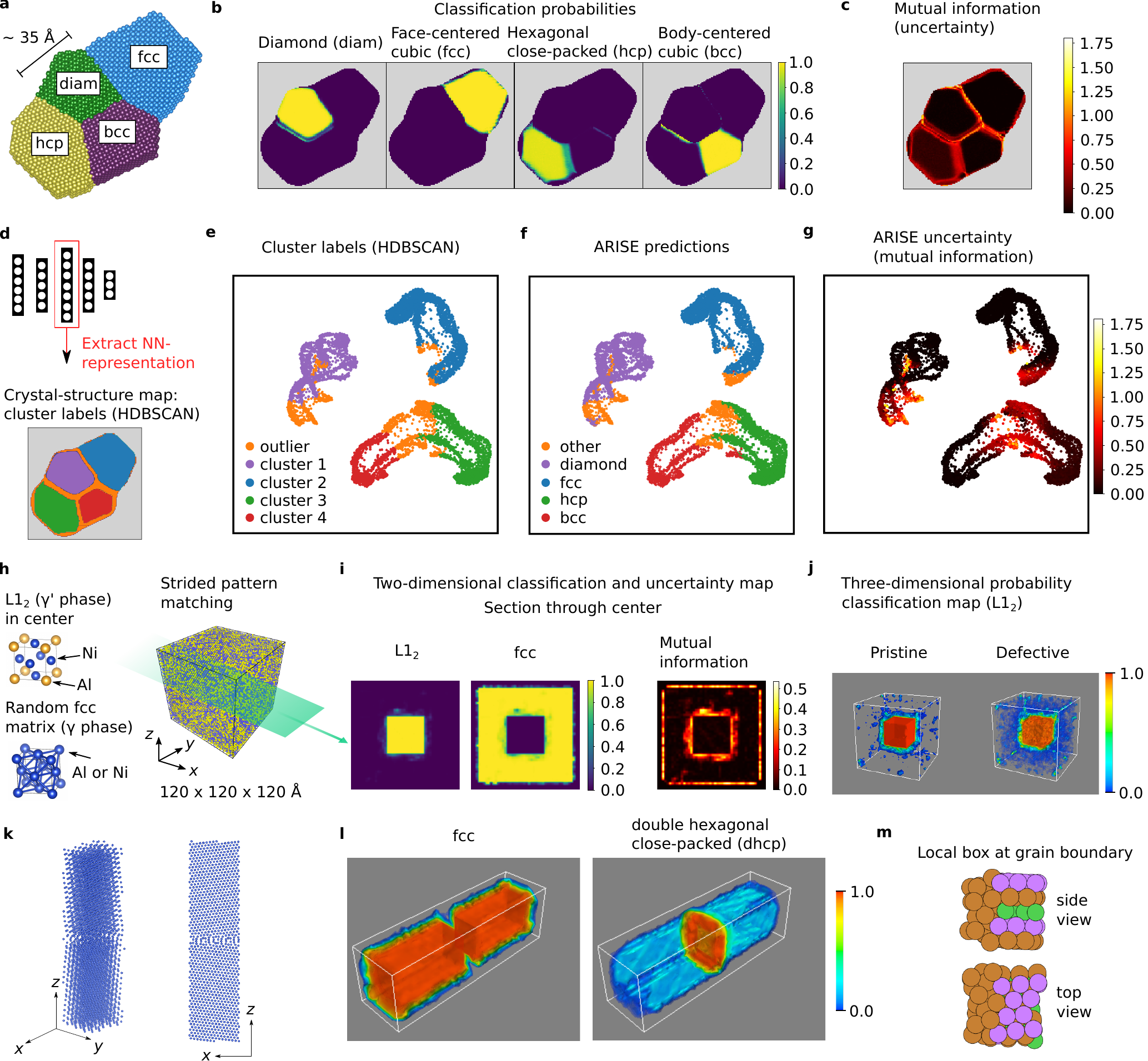}
  \caption{\textbf{Analysis of synthetic polycrystals.} \textbf{a} Mono-species polycrystal consisting of four grains with  face-centered cubic (fcc), body-centered cubic (bcc),
  hexagonal close-packed (hcp), and diamond (dia) symmetry. 
  \textbf{b} Classification probabilities of expected prototypes.  
  \textbf{c} Mutual information map for uncertainty quantification. 
  \textbf{d-g} Unsupervised analysis of internal neural-network representations.  
  \textbf{d} The neural-network representations are extracted for each local segment in \textbf{a} (obtained via SPM). Clustering (via Hierarchical Density-based Spatial Clustering Applications with Noise, HDBSCAN)
  is applied to this
  high-dimensional 
  space; 
  the polycrystal is marked according to the resulting clusters (see legend in \textbf{e} for the color assignments). 
 \textbf{e-g} Two-dimensional projection (via Uniform Manifold Approximation and Projection, UMAP) of neural-network representations colored by cluster label,  
ARISE predicted class,  
 and mutual information, respectively. 
 In \textbf{e}, all points for which HDBSCAN does not assign a cluster are labeled as outlier. In \textbf{f}, all 
 points that are not classified as fcc, diamond, hcp or bcc are labeled as other. 
 Note that while the distances 
 between points are meaningful, the axes merely serve as a bounding window and are not interpretable \textendash\ a situation typically encountered in non-linear methods such as UMAP (cf. section 6\cite{mcinnes2018umap}).
 \textbf{h-j} Precipitate detection in Ni-based superalloys. 
 \textbf{h} Binary model system (right) and depiction of the two appearing phases (left). 
 \textbf{i} Classification probabilities of expected prototypes and mutual information for a slice through the center of the structure. 
 \textbf{j} 3D-resolved detection of the precipitate via the L1$_2$ classification probability for the pristine (left) and highly-defective case (right), for which 20\% of the atoms are removed and randomly displaced (up to 5\% of the nearest neighbor distance). 
\textbf{k} Lowest-energy grain boundary structure (Cu, fcc) predicted from an evolutionary search. The so-called Pearl pattern appears at the grain boundary, which is also observed in experiment\cite{meiners2020observations}. 
\textbf{l} SPM-ARISE analysis, correctly identifying fcc (ABC close-packing) in the grains, while detecting double hexagonal close-packed (dhcp, ABAC) at the grain boundary.
\textbf{m} Exemplary analysis of a local box at the grain boundary, illustrating a change in stacking and increased distortions, which motivates the assignment of dhcp (which contains 50\,\% of both fcc and hcp close-packings).
 }
  \label{fig:synthetic_polycrystal}
 \end{figure*}

\begin{figure*}[!htbp]
 \includegraphics[width=\textwidth]{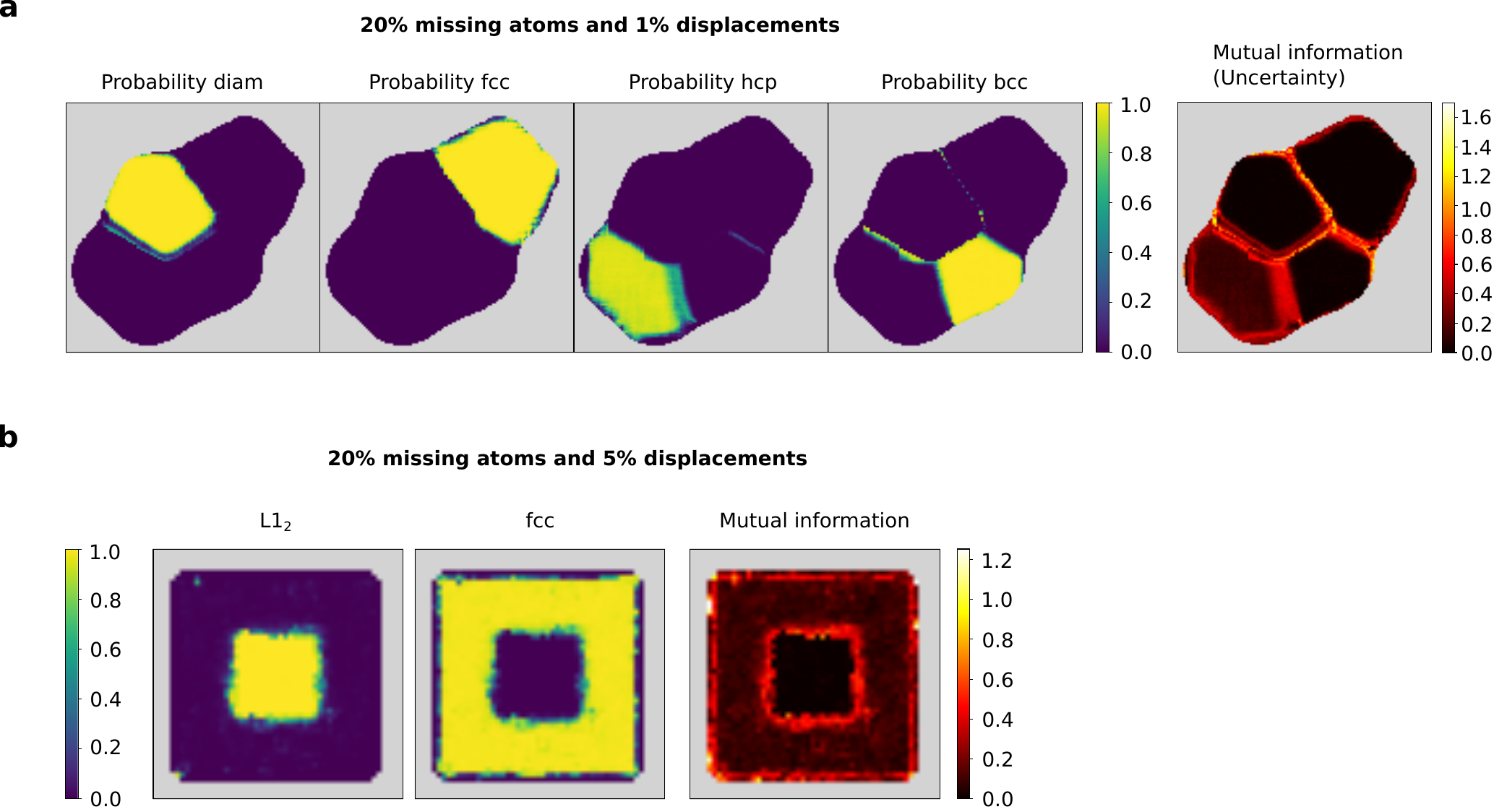}
 \caption{Investigation of distorted synthetic polycrystals. \textbf{a} Analysis on defective version of mono-species polycrystal shown in Fig. \ref{fig:synthetic_polycrystal}a. \textbf{b} Analysis on defective version 
 of superalloy structure shown in Fig. \ref{fig:synthetic_polycrystal}h. \label{fig:supp_four_grains_defective}}
\end{figure*}

First, the classification via SPM combined with ARISE is performed for a slab-like synthetic polycrystal consisting of fcc, bcc, hcp, and diamond grains (cf. Fig. \ref{fig:synthetic_polycrystal}a). 
Due to the nature of the system, the SPM boxes near the
grain boundaries   
will contain mixtures of different crystal structures. 
The results are shown in Fig. \ref{fig:synthetic_polycrystal} b and c:
The network assigns high classification probability 
to the correct prototypes.
Uncertainty is low within the grains, increasing at grain boundaries and crystal outer borders in line with physical intuition.
The result remains virtually unchanged  when introducing 
atomic displacements (up to $1\%$ of the nearest neighbor distance) while concurrently removing 20\% of the atoms (cf. Supplementary Fig. \ref{fig:supp_four_grains_defective}). 
The highest classification probabilities (after from the top four shown in Fig. \ref{fig:synthetic_polycrystal}b) are shown in Supplementary Fig. \ref{fig:suppl_four_grains}; a discussion on the stride can be found in Supplementary Fig. \ref{fig:low_resolution_four_grains}.

Going beyond classification, we show how unsupervised learning can be used to access structural similarity information embedded in ARISE's internal representations, and use it for the characterization of crystal systems.
We consider the mono-species polycrystal shown in Fig. \ref{fig:synthetic_polycrystal}a  
and collect ARISE's representations of the overall 7$\,$968 local boxes. 
Next, we employ Hierarchical Density-based Spatial Clustering Applications with Noise (HDBSCAN)\cite{mcinnes2017accelerated, McInnes2017} to identify clusters in the high-dimensional representation space. 
HDBSCAN estimates the density underlying a given data set and then constructs a hierarchy of clusters, from which the final clustering can be obtained via an intuitive and tunable parameter (see Methods). 
The obtained clusters  correspond to the four crystalline grains in the structure (Fig. \ref{fig:synthetic_polycrystal}d). 
Points identified as outliers (marked in orange) coincide with grain-boundary and outer-border regions. 
Next, the high-dimensional manifold of the NN representations is projected in 2D via Uniform Manifold Approximation and Projection (UMAP)\cite{mcinnes2018umap}. 
 UMAP models the manifold underlying a given dataset and then finds a low-dimensional projection that can capture both global and local distances of the original high-dimensional data. 
This returns a structure-similarity map (Fig. \ref{fig:synthetic_polycrystal}e), which allows to visually investigate similarities among structures: 
points (structures) close to each other in this map are considered to be similar by the algorithm.
Structures belonging to the same cluster are in close proximity to each other, and clearly separated from other clusters. 
Conversely, outlier points are split across different regions of the map. 
This is physically meaningful: outliers are not a cohesive cluster of similar structures, but rather comprise different types of grain boundaries 
(i.e., fcc to bcc transitions or fcc to diamond etc., cf. Supplementary Fig. \ref{fig:supp_hdbscan_pos_gb}). 
In this synthetic setting, we can also use the classification prediction to further verify the unsupervised analysis: 
the results obtained via unsupervised learning indeed match ARISE's predictions (cf. Fig. \ref{fig:synthetic_polycrystal}e - Fig \ref{fig:synthetic_polycrystal}f).
Moreover, an analysis of the mutual information (Fig. \ref{fig:synthetic_polycrystal}g) reveals that points at the core of the clusters are associated with low uncertainty, while points closer to the boundaries show increased uncertainty. 
Similar results are obtained for the other layers (cf. Supplementary Fig. \ref{fig:four_grains_umap_hdbscan_full}). 

We now move to a more realistic system: a model structure for Ni-based superalloys\cite{reed2008superalloys} (c.f Fig. \ref{fig:synthetic_polycrystal}h). 
Ni-based superalloys are used in aircraft engines due to their large mechanical strength at high temperatures, which derives 
from ordered L$1_2$ precipitates ($\gamma^\prime$ phase) embedded in a fcc matrix ($\gamma$ phase). 
We generate an atomic structure consisting of a fcc matrix in which Al and Ni atoms are randomly distributed. 
In the center, however, the arrangement of Al and Ni atoms is no longer random, but it is ordered such that the L$1_2$ phase is created (c.f Fig. \ref{fig:synthetic_polycrystal}h). 
The cubic shape of this precipitate is in accordance with experimental observations\cite{raabe_exp_superalloy}. 
The resulting structure 
comprises 132$\,$127 atoms over a cube of $120\,\text{\AA}$ length. 
As shown via a section through the center in Fig. \ref{fig:synthetic_polycrystal}i, fcc is correctly assigned to the matrix, and the precipitate is also detected. 
The uncertainty is increased at the boundary between random matrix and precipitate, as well as at the outer borders. 
Fig. \ref{fig:synthetic_polycrystal}j illustrates the L1$_2$ classification probability in a 3D plot. The precipitate is detected in both pristine and highly-defective structures. 
This analysis demonstrates that ARISE can distinguish between chemically ordered and chemically disordered structures, a feature that will be exploited for the analysis of experimental data in Sec. Application 
to atomic-electron-tomography data.

\begin{figure*}[!htbp]
 \includegraphics[width=\textwidth]{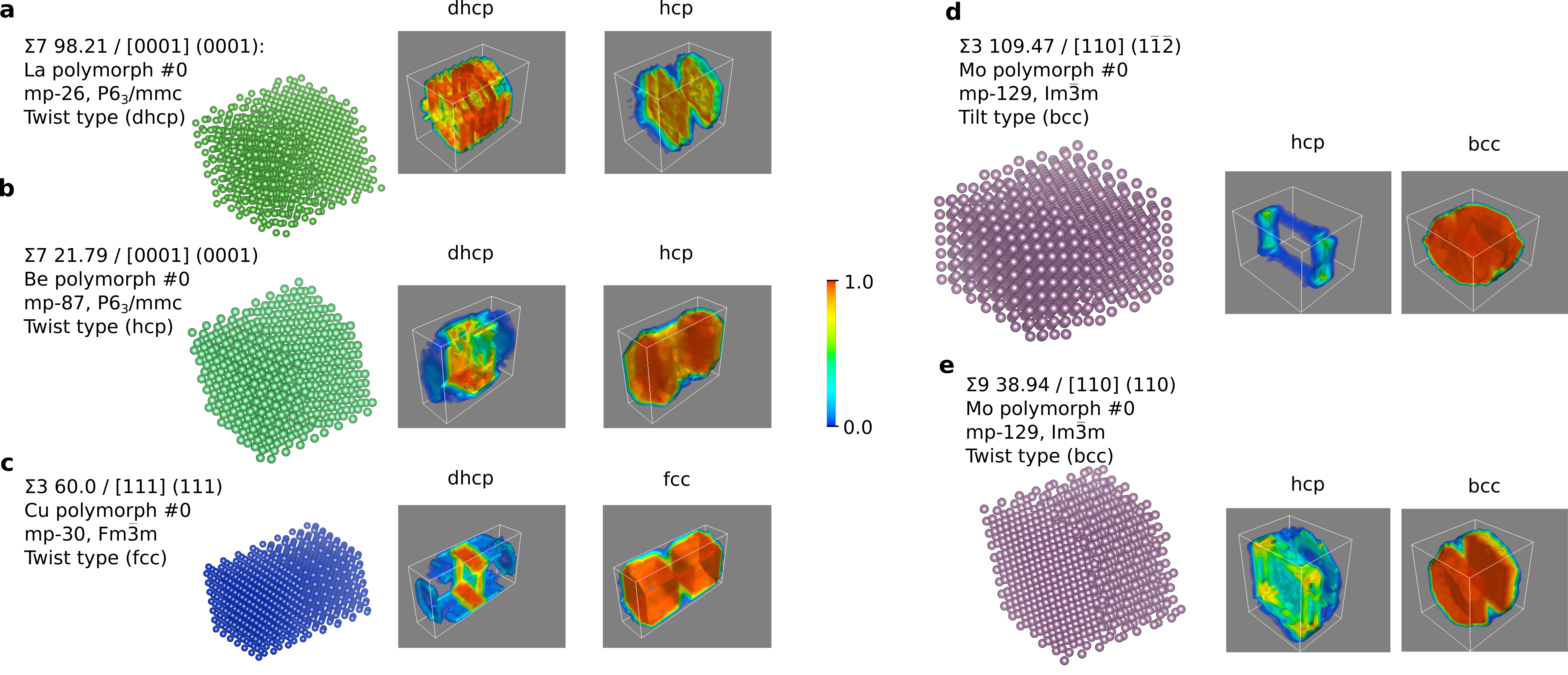}
 \caption{Five representative examples from the grain boundary database (GBDB), which is the largest, currently available database of DFT-computed grain boundary properties\cite{zheng2020grain}. 
 This database can be accessed via Materials Project or \url{http://crystalium.materialsvirtuallab.org/}.  
 For each structure, four lines of information are provided:
 The first line specifies the information that is required to uniquely describe a grain boundary structure\cite{lejvcek2010grain}, where 
 first the $\Sigma$-parameter 
 is given, followed by rotation angle, rotation axis and grain-boundary plane. 
 The relative orientation of two neighboring grains is described by three degrees of freedom (rotation angle and axis). 
 The two degrees of freedom specified via the grain-boundary plane complete the unique characterization of a grain-boundary structure.    
  The second line specifies the element and the entry number of the polymorph in the database (for a given element, multiple grain boundaries can be available). 
 The third line specifies the materials project ID and the space group. The last line specifies the grain-boundary type (twist, tilt) alongside the dominating crystal structure. 
 The database entries correspond to periodic cells that contain a grain boundary. We replicate this initial cell isotropically (in the plane parallel to the grain boundary) until at least 1000 atoms are contained in the structure.
For all examples, the dominating phase and grain boundary regions are correctly detected as shown via the 3D classification probability maps 
of the most popular assignments according to ARISE. These selected structures illustrate the advantages of ARISE in the following way: 
\textbf{a} shows that ARISE can detect dhcp symmetry in a polycrystal. In particular, the close-packing corresponding to dhcp cannot be classified in comparable automatic 
fashion by any of the available methods. 
For hcp (\textbf{b}) and fcc (\textbf{c}), the dhcp assignments only appear at the grain boundary. \textbf{d} and \textbf{e}  are two 
different grain boundary types that do not only differ in their defining degrees of freedom but also are of tilt (\textbf{d}) and twist (\textbf{e}) type. 
ARISE distinguishes the local structures at the grain boundary which is indicated by its assignments: while for the twist type (\textbf{e}) 
hcp is the dominating assignment at the grain boundary, for the tilt type the hcp probability drops to zero at the grain boundary (except for the outer borders). 
The following SPM parameters are chosen for all examples: A stride of 2\,\AA\ suffices to resolve the main characteristics. For a box size of 16\,\AA\ at least 100 atoms are contained in the boxes within the grains. 
 \label{fig:supp_gb_database}}
\end{figure*}

 Another realistic  
system is shown in Fig. \ref{fig:synthetic_polycrystal}k, which 
 is the lowest-energy structure obtained from an evolutionary structure search\cite{meiners2020observations}. 
The structural patterns at the grain boundary are also observed in scanning transmission electron microscopy (STEM) experiments. 
SPM-ARISE correctly identifies the fcc symmetry within the grains (cf. Fig. \ref{fig:synthetic_polycrystal}l) while assigning double hexagonal close-packed (dhcp) symmetry at the grain boundary. 
The local boxes at the grain boundary 
 contain partial fcc structures 
while changes in stacking and distortions decrease their symmetry (cf. Fig. \ref{fig:synthetic_polycrystal}m).  
Also the dhcp phase (ABAC close-packing) contains fcc (ABC) and a lower-symmetry packing (hcp, AB), thus justifying the assignment.  
To supplement this study, we investigate several examples from the largest, currently available grain-boundary database\cite{zheng2020grain}, including fcc, bcc, hcp, and dhcp symmetry 
as well as various grain 
boundary types, which ARISE can classify correctly (cf. Supplementary Figure \ref{fig:supp_gb_database}). Note that ARISE correctly identifies even 
the $\alpha-$Sm-type stacking (ABCBCACAB).  
No other fully automatic approach offers a comparable sensitivity. 

\subsection*{Application to transmission-electron-microscopy experimental images}

\begin{figure*}
   \begin{center}
     \includegraphics[width=\textwidth]{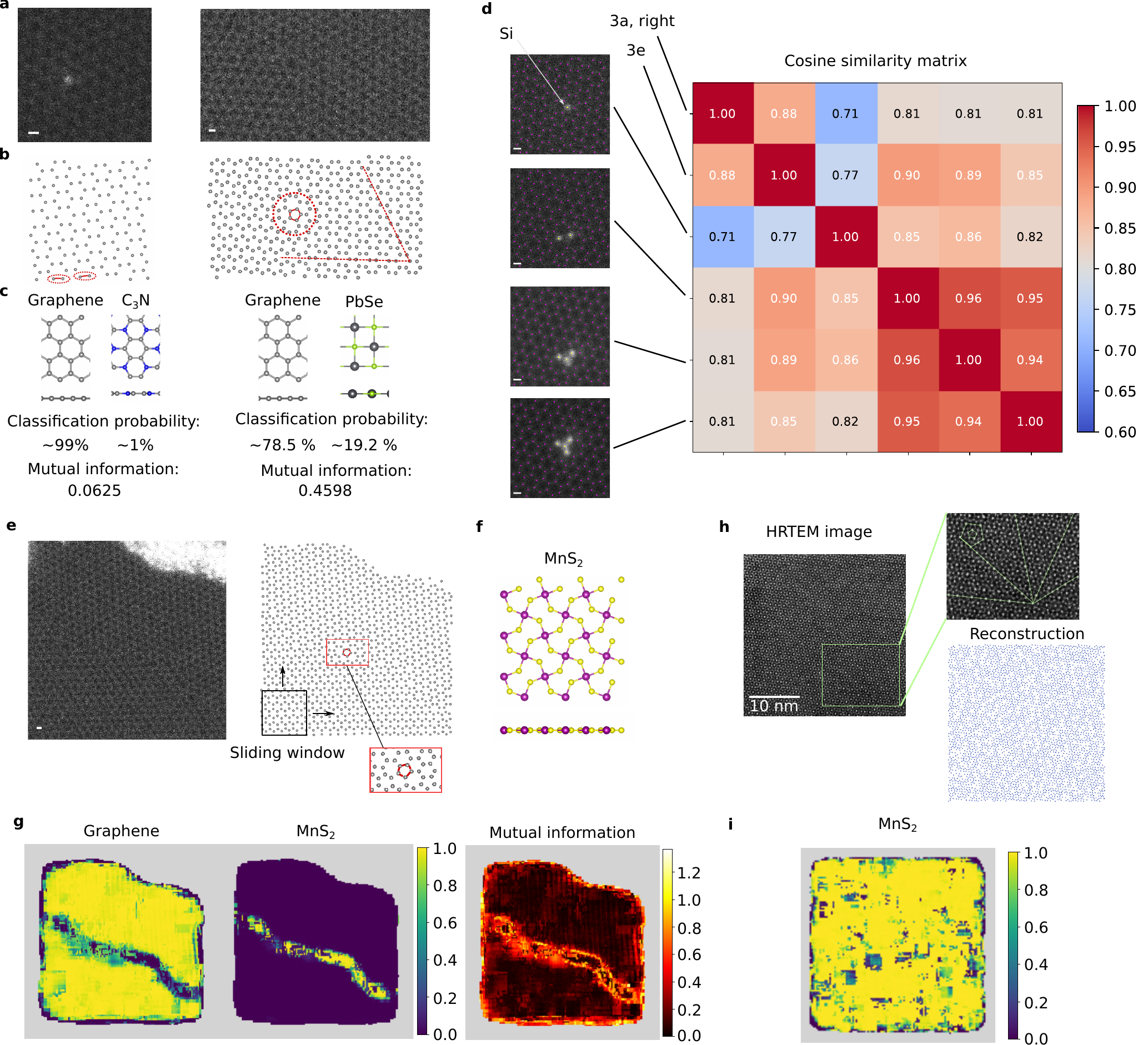}
   \end{center}
   \caption{\textbf{Analysis of HAADF and HRTEM images via ARISE and SPM.} 
   \textbf{a} Experimental high-angle annular dark field (HAADF) images of two graphene structures. White scale bars in all HAADF images in this figure are positioned in the bottom left and correspond to the typical graphene bond length (1.42\,\AA).  
   \textbf{b} The atomic positions are reconstructed from the images via AtomNet\cite{ziatdinov2017deep}. 
   \textbf{c} The resulting atomic structures are analyzed using ARISE. 
    The top predicted structures are shown. Mutual information is used to quantify the classification uncertainty.
    \textbf{d} Similarity quantification of HAADF images via ARISE. The images in \textbf{a} (right) and \textbf{e} are compared to a selection 
    of graphene systems with Si defects\cite{ziatdinov2019building}. For each image, AtomNet is used for reconstruction and the internal  representations of ARISE are extracted (here, second hidden layer).
    Then, the cross-similarity is calculated using the cosine similarity. A block matrix structure arises that correlates 
    with the number of Si atoms. A similar pattern is observed for a larger selection of structures, cf. Supplementary Fig. \ref{fig:supp_STEM_sim}.  
   \textbf{e} HAADF image and reconstructed atomic positions (analogous to \textbf{a-b}) of a larger sample. Pentagons can be spotted near the grain boundary (see inset). 
   \textbf{f} MnS$_2$ prototype. 
   \textbf{g} Local analysis via strided pattern matching: graphene is the dominant structure. Different prototypes (MnS$_2$) are only assigned - and with high uncertainty (mutual information) - at the grain boundary.
   \textbf{h} High resolution transmission electron microscopy (HTREM) image of a quasicrystalline structure (icosahedral Al-Cu-Fe, adapted from the original reference\cite{li2016review}, see Methods). While there is an underlying order, the structure 
   is aperiodic (i.e., no translational symmetry is present). As visualized in the zoom, the bright spots align with five-fold symmetry axes and pentagons of different size appear. 
   Based on the reconstruction via AtomNet (bottom right), ARISE (via strided pattern matching) identifies MnS$_2$ as the 
   dominating prototype (\textbf{i}), which similarly to the input structure contains pentagon patterns (\textbf{f}). 
   }
   \label{fig:STEM}
 \end{figure*}

We now investigate defective structures originating from a completely different data source,  
namely STEM experiments, 
to demonstrate the generalization ability of ARISE and its applicability to experimental data. 
Moreover, we show how global and local analysis can be combined to analyze crystal structures. 
STEM experiments are a valuable resource to characterize material specimens, and to study, for instance, the atomic structures at grain boundaries\cite{meiners2020observations}. 
Atomic resolution can be achieved in high-angle annular dark field (HAADF) images. 
The global assignments of ARISE are tested on two experimental HAADF images of graphene shown in Fig. \ref{fig:STEM}a. 
These images contain a substantial amount of noise which makes it very challenging to recognize the graphene honeycomb pattern by naked eye. 
The choice of graphene is motivated by it being a flat 2D materials; $x$ and $y$ atomic positions obtained from STEM images thus provide the actual crystal structure, and not a mere projection. 
Approximate atomic positions (i.e. $x$ and $y$ coordinates) from HAADF images are obtained via AtomNet\cite{ziatdinov2017deep}, and shown in Fig. \ref{fig:STEM}b.
ARISE is then used to classify the structures following the steps summarized in Fig. \ref{fig:single_and_polyc_class_steps}a-d.
The top predictions ranked by classification probability are shown in Fig. \ref{fig:STEM}c, together with the uncertainty of the assignments as quantified by the mutual information.
ARISE correctly recognizes both images as graphene, despite the substantial amount of noise present in images and reconstructed atomic positions. 
For the first image (Fig. \ref{fig:STEM}a, left), graphene is predicted with very high probability ($\sim 99\%$). 
Indeed, the similarity to graphene is apparent, although evident distortions are present in some regions (e.g., misaligned bonds marked in Fig. \ref{fig:STEM}b). 
The second candidate structure is C$_3$N, predicted with $\sim 1\%$ probability; in C$_3$N, atoms are arranged in a honeycomb lattice, making also this low probability assignment physically meaningful. 
For the second image (Fig. \ref{fig:STEM}a, right), ARISE also correctly predicts graphene, this time with 79$\%$ probability. The uncertainty is six times larger than in the previous case.
Indeed, this structure is much more defective than the previous one: it contains a grain boundary in the lower part, causing evident deviations from the pristine graphene lattice, as illustrated in Fig. \ref{fig:STEM}b (right). 
The other four candidate structures appearing in the top five predictions (PbSe, MnS$_2$, BN, C$_3$N) are the remaining completely flat monolayers known to the network (out of the 108 structures in the training dataset, only five are flat monolayers). 
Note that no explicit information about the dimensionality of the material is given to the model. 
It is also important to point out that ARISE robustness well beyond physical levels of noise is essential to achieve the correct classification despite 
the presence of substantial amount of noise from both experiment and atomic position reconstruction. 

\begin{figure*}[!htbp]
 \includegraphics[width=\textwidth]{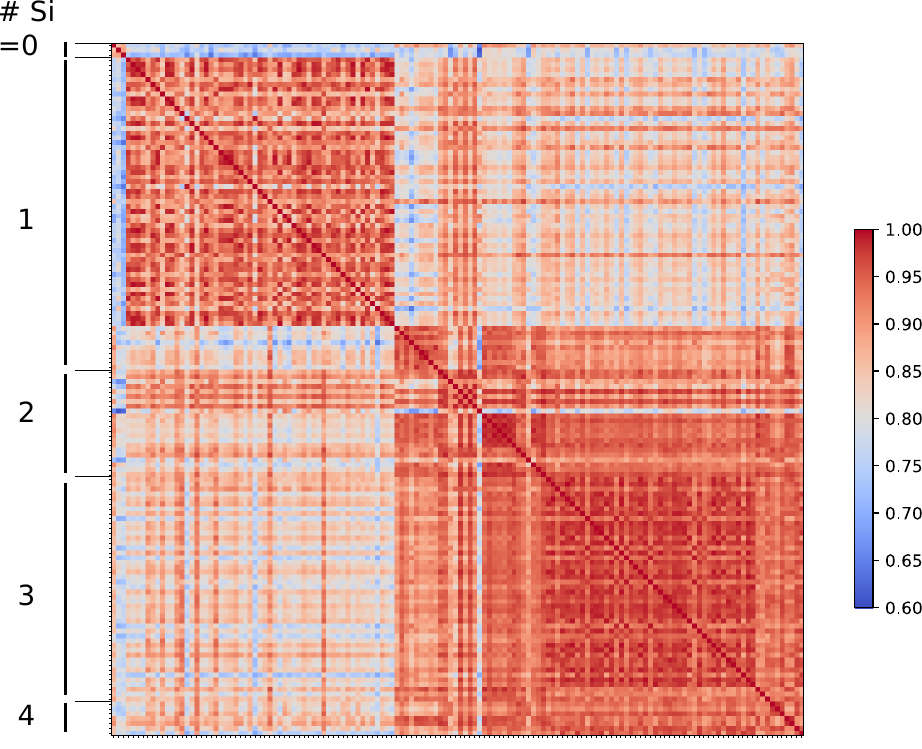}
 \caption{Cross similarity matrix for a selection of the defect library\cite{ziatdinov2019building} that is larger than in the main text (Fig. \ref{fig:STEM}d). Specifically, 140 structures as well as  
 the mono-species structures from Fig. \ref{fig:STEM}a (right),e are considered. For reconstruction of atomic positions, Atomnet is employed, 
 where for the structures from the library, atomic positions are reconstructed using a model that can also classify the chemical species. We employed 
 the model that is available at \url{https://github.com/pycroscopy/AICrystallographer/tree/master/DefectNet}. 
 \label{fig:supp_STEM_sim}}
\end{figure*}

Besides the separate classification of single images,  
ARISE also learns meaningful similarities between images  (i.e. structures). 
To demonstrate this, we analyze a library of graphene images with Si defects\cite{ziatdinov2019building} and quantify their similarity using ARISE's internal representations. 
Fig. \ref{fig:STEM}d investigates a selection of images which contain the mono-species structures of Fig. \ref{fig:STEM}a (right), e, and systems with up to four Si atoms. 
Atomic positions are determined via AtomNet. Then, the 
internal representations 
from ARISE are extracted and the pairwise cosine similarity is calculated.
The cross-similarity matrix is depicted in Fig. \ref{fig:STEM}d, revealing a block matrix form in which the  binary and mono-species structures are separated, i.e., more similar to each other, which 
can be attributed to the number of Si defects.
This characteristic reappears for a larger selection of structures (cf. Supplementary Fig. \ref{fig:supp_STEM_sim}), thus confirming the analysis 
of Fig. \ref{fig:STEM}d. This investigation demonstrates that ARISE learns meaningful similarities, supporting the general applicability of ARISE for similarity quantification. 

 While so far we have analyzed HAADF images on a global scale, a local analysis via SPM allows to zoom into a given structure and locate sub-structural features. 
This is particularly useful for polycrystalline and/or larger systems (e.g., more than 1$\,$000 atoms).  
As illustrative example, we consider the structure in Fig. \ref{fig:STEM}e. 
The mutual information shown in Fig. \ref{fig:STEM}g (right) clearly reveals the presence of a grain boundary. 
In Fig. \ref{fig:STEM}g (left), the classification probabilities of graphene and MnS$_2$ (the dominant prototypes) are presented, the latter being assigned at the grain boundary. 
This assignment can be traced back to pentagon-like patterns appearing near the grain boundary (as highlighted in Fig. \ref{fig:STEM}e), a pattern similar to the one being formed by Mn and S atoms in MnS$_2$ (cf. Fig. \ref{fig:STEM}f). 

Next, we challenge the established procedure for the local analysis of 2D images with data from a completely different resource. 
We investigate a high-resolution transmission electron microscopy (HTREM) image of a quasicrystalline structure\cite{levine1984quasicrystals, li2016review}, cf. Fig \ref{fig:STEM}h. 
The bright spots are ordered aperiodically, making it a very hard task to identify the underlying order by eye. 
Via the established procedure, MnS$_2$ is predicted as the most similar prototype (cf. Fig. \ref{fig:STEM}i). MnS$_2$ contains pentagon patterns (cf. Fig. \ref{fig:STEM}f) which 
 can also be seen in the quasicrystal (cf. zoom in Fig. \ref{fig:STEM}h). This result suggests that ARISE and SPM are novel and promising tools for the classification of 
quasicrystalline order in automatic fashion \textendash\  a promising yet under-explored area.  

\subsection*{Application to atomic-electron-tomography data}
While HAADF images are a valuable experimental resource, they only provide 2D projections.
3D structural and chemical information can however be obtained from atomic electron tomography (AET) 
with atomic resolution achieved in recent experiments\cite{miao2016atomic, zhou2020atomic, chen2013three, xu2015three}.  
Notably, this technique provides 3D atomic positions 
labeled by chemical species, to which ARISE and SPM can be readily applied. 
While extensions to other systems such as 2D materials are reported\cite{tian2020correlating}, metallic nanoparticles have been the main experimental focus so far, 
specifically FePt systems due to their promises for biomedicine and magnetic data storage\cite{sun2006recent}. 
First, a FePt nanoparticle\cite{yang2017deciphering} is classified using SPM-ARISE. 
ARISE's robustness is critical for this application, since the structural information provided by AET experiments are based on reconstruction algorithms 
that cause visible distortions (cf. Fig. \ref{fig:AET}a). 
SPM-ARISE primarily detects L1$_2$, L1$_0$, and fcc phases (see Supplementary Fig. \ref{fig:supp_nanoparticle_2017_uncertainty}).
This is in line with physical expectations: annealing leads to structural transitions from chemically disordered to ordered fcc (A1 to L1$_2$) or to the tetragonal L1$_0$ phase\cite{sun2006recent, yang2017deciphering}. 
Besides the expected prototypes, ARISE also finds regions similar to tetragonally distorted, mono-species fcc (cf. Supplementary Fig. \ref{fig:supp_nanoparticle_2017_uncertainty}), which is meaningful given the presence of fcc and the tetragonal phase L1$_0$. 

To go beyond the information provided by classification and discover hidden patterns and trends in AET data, we conduct an exploratory analysis using unsupervised learning on ARISE's internal representations. 
While the procedure is similar to the one presented in Fig. \ref{fig:synthetic_polycrystal}d-g, here the analysis is truly exploratory (no ground truth is known), and data comes from experiment. 
First, all SPM boxes classified as L1$_0$ are extracted, this choice motivated by the physical relevance of this phase, in particular due to its magnetic properties\cite{sun2006recent}. 
This reduces the number of data points (boxes) from 43$\,$679 to 5$\,$359 \textendash\ a significant filtering step for which the automatic nature of ARISE is essential. 
In the representation space of the first hidden layer, HDBSCAN identifies seven clusters (and the outliers). 
To interpret the cluster assignments, we analyze geometrical characteristics of atomic structures (i.e., the local boxes) assigned to the different clusters. 
Specifically, we consider the nearest neighbor distances between Fe and Pt atoms, $d_{\text{FeFe}}$ and $d_{\text{PtPt}}$, respectively 
(cf. Supplementary Methods  
for the definition). 
For an ideal tetragonal structure, the difference $\Delta d = d_{\text{FeFe}} - d_{\text{PtPt}}$ is zero (cf. Fig. \ref{fig:AET} b, top left); a deviation from this value thus quantifies the level of distortion. 
Looking at the histograms of the (signed) quantity $\Delta d$ shown in  Fig. \ref{fig:AET}b for each cluster, one can observe that 
each distribution is peaked; moreover, the distribution centers vary from negative to positive $\Delta d$ values across different clusters. The distribution of the outliers is shown for comparison: the $\Delta d$ distribution is very broad, 
since outlier points are not a meaningful cluster. 
While overlap exists, the clusters correspond to subgroups of structures, each distorted in a different way, as quantified by $\Delta d$. 
Thus, we discovered a clear trend via the cluster assignment that correlates with the level of distortion. 
The cluster separation can be visualized in 2D via UMAP (cf. Fig. \ref{fig:AET}b).
Notably, the clusters do not overlap, even in this highly compressed representation (from 256 to 2 dimensions).
Some of the clusters may also contain further sub-distributions, which seems apparent for instance from the $\Delta d$ distribution of cluster 6.  
The regions corresponding to the clusters could be hinting at a specific growth mechanism of the L1$_0$ phase during annealing, although further investigations are necessary to support this claim. 
The present analysis provides a protocol for the machine-learning driven exploration of structural data: supervised learning is employed to filter out a class of interest 
(which is not a necessary step, cf. Fig. \ref{fig:synthetic_polycrystal}d-g), then unsupervised learning is applied to the NN representations, revealing regions sharing physically meaningful geometrical characteristics. 

Finally, we apply ARISE to time-resolved (i.e., four-dimensional) AET data. Specifically, a nanoparticle measured for three different annealing times is investigated\cite{zhou2019observing}. 
The mutual information as obtained via SPM-ARISE  
is shown in Fig. \ref{fig:AET}c for five central slices. 
In regions between outer shell and inner core, the mutual information clearly decreases for larger annealing times, indicating that crystalline order increases inside the nanoparticle (see also Supplementary Fig. \ref{fig:AET_annealing_supp} for more details).
This analysis confirms that the predictive uncertainty of ARISE, as quantified by the mutual information, directly correlates with crystalline order. 
The mutual information can be therefore considered an AI-based order parameter, which we anticipate to be useful in future nucleation dynamics studies.  

 \begin{figure*}
   \centering
     \includegraphics[width=\textwidth]{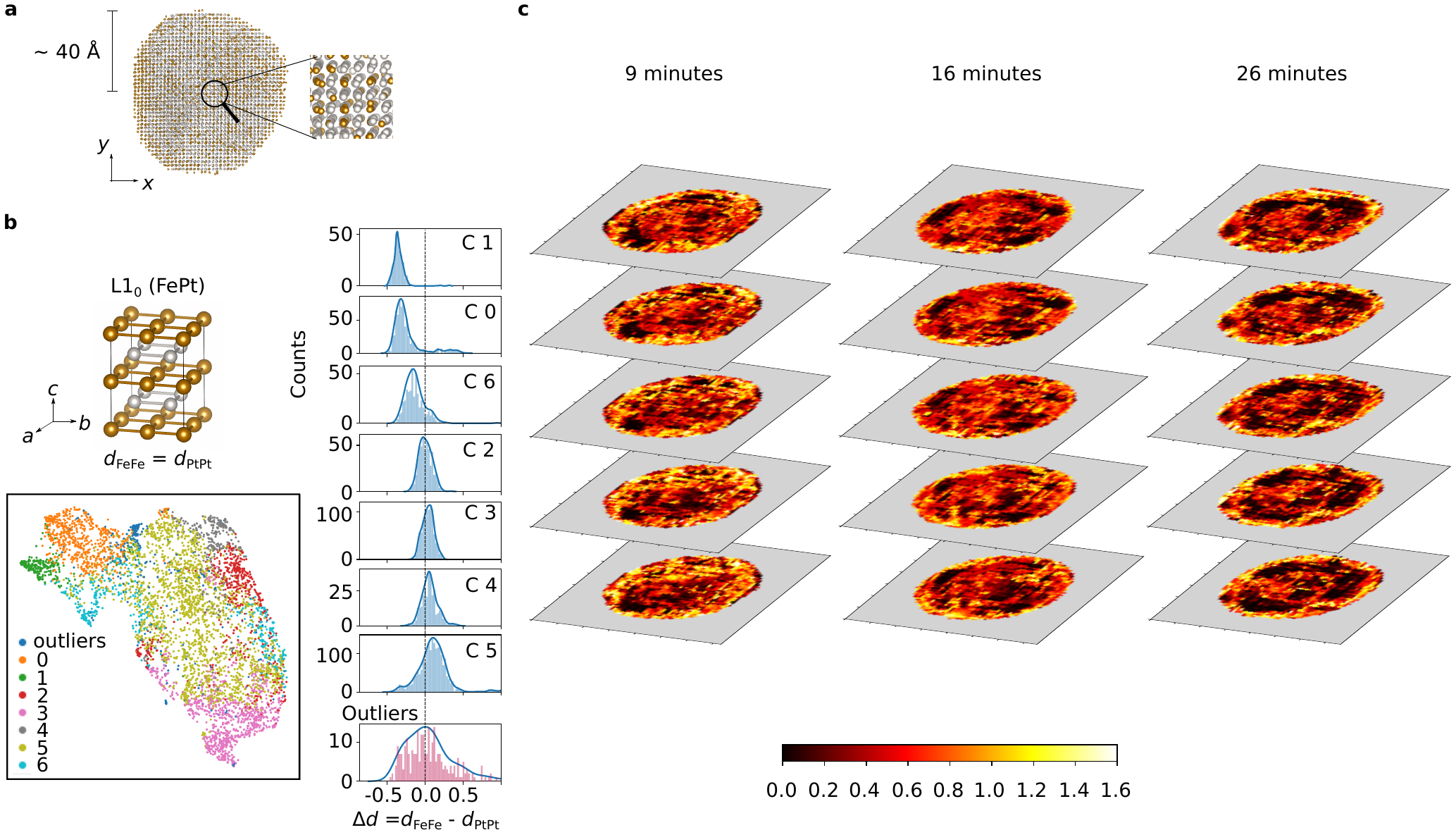}
    \caption{\textbf{Analysis of atomic electron tomography data.} \label{fig:AET}
    \textbf{a} Side view of FePt nanoparticle ($\sim$23k atoms),  
    with atomic positions and chemical species from atomic electron tomography (AET) data\cite{yang2017deciphering}. 
\textbf{b} Two-dimensional projection (bottom left) of neural-network representations (first hidden layer) via UMAP for regions classified as L1$_0$ by ARISE. 
The distribution of the difference between the nearest neighbor distances $d_{\text{FeFe}}$ and $d_{\text{PtPt}}$ (highlighted by bonds in top left part) is shown for each cluster (right), where 
 cluster $i= 0, ..., 6$ is denoted as Ci, while all points for which HDBSCAN does not assign a cluster are labeled as outlier.  
    \textbf{c} Five central slices (mutual information, obtained via strided pattern matching) for three different annealing times (data from four-dimensional AET experiment\cite{zhou2019observing}). 
    }
  \end{figure*}

\section*{Discussion}
In this work, Bayesian deep learning is employed to achieve a flexible, robust and threshold-independent crystal classification model, which we term ARISE. 
This approach correctly classifies a comprehensive and diverse set of crystal structures from computations and experiments, including polycrystalline systems (via strided pattern matching). 
 Given an unknown structure, the network assigns \textendash\ in an automatic fashion \textendash\ the most similar prototypes among 108 possible classes (and quantifies the similarity!), which 
 is a very complicated task even for trained materials scientists, in particular in case of complex and possibly defective 3D structures. 
 ARISE is trained on ideal synthetic systems only and correctly identifies crystal structures in STEM and AET experiments, hence demonstrating strong generalization capabilities. 
The Bayesian deep-learning model provides classification probabilities, which - at variance with standard NNs - allow for the quantification of predictive uncertainty via mutual information. 
The mutual information is found to directly correlate with the degree of crystalline order, as shown by the analysis of time-resolved data from AET experiments. 
This demonstrates the correlation of an information-theory concept with physical intuition. 
The internal NN representations are analyzed via state-of-the-art unsupervised learning. 
The clusters identified in this high-dimensional internal space allow to uncover physically meaningful structural regions. 
These can be grain boundaries, but also unexpected substructures sharing geometrical properties, as shown for metallic nanoparticles from AET experiments. 
This illustrates how supervised and unsupervised machine learning can be combined to discover hidden patterns in materials science data. 
In particular, the physical content learned by the NN model is explained by means of unsupervised learning. 
Since ARISE is not limited to predicting the space group, systems where the space group does not characterize the crystal structure can be tackled (as demonstrated for carbon nanotubes). 
More complex systems such as
quasi-crystals\cite{levine1984quasicrystals},
periodic knots, or weavings\cite{liu2018geometry} could also be considered. 
Indeed, ARISE can be applied to any data 
providing Cartesian coordinates labeled by chemical species.
Practically, one simply needs to add the new structures of interest to the training set, and re-train or fine-tune (i.e., via transfer learning) the NN with the desired labels. 
Moreover, the mutual information allows to 
quantify the defectiveness of a structure; this could be exploited to automatically evaluate the quality of STEM images, for example one may automatically screen for STEM images that are likely to contain structural defects.
Applications in active learning\cite{gal2017deep} for materials science are also envisioned, where uncertainty is crucial for example when deciding on the inclusion of additional - typically computationally costly - points in the dataset.

 \clearpage

\section*{Methods}

\begin{figure*}[!htb]
\centering
\includegraphics[width=\textwidth]{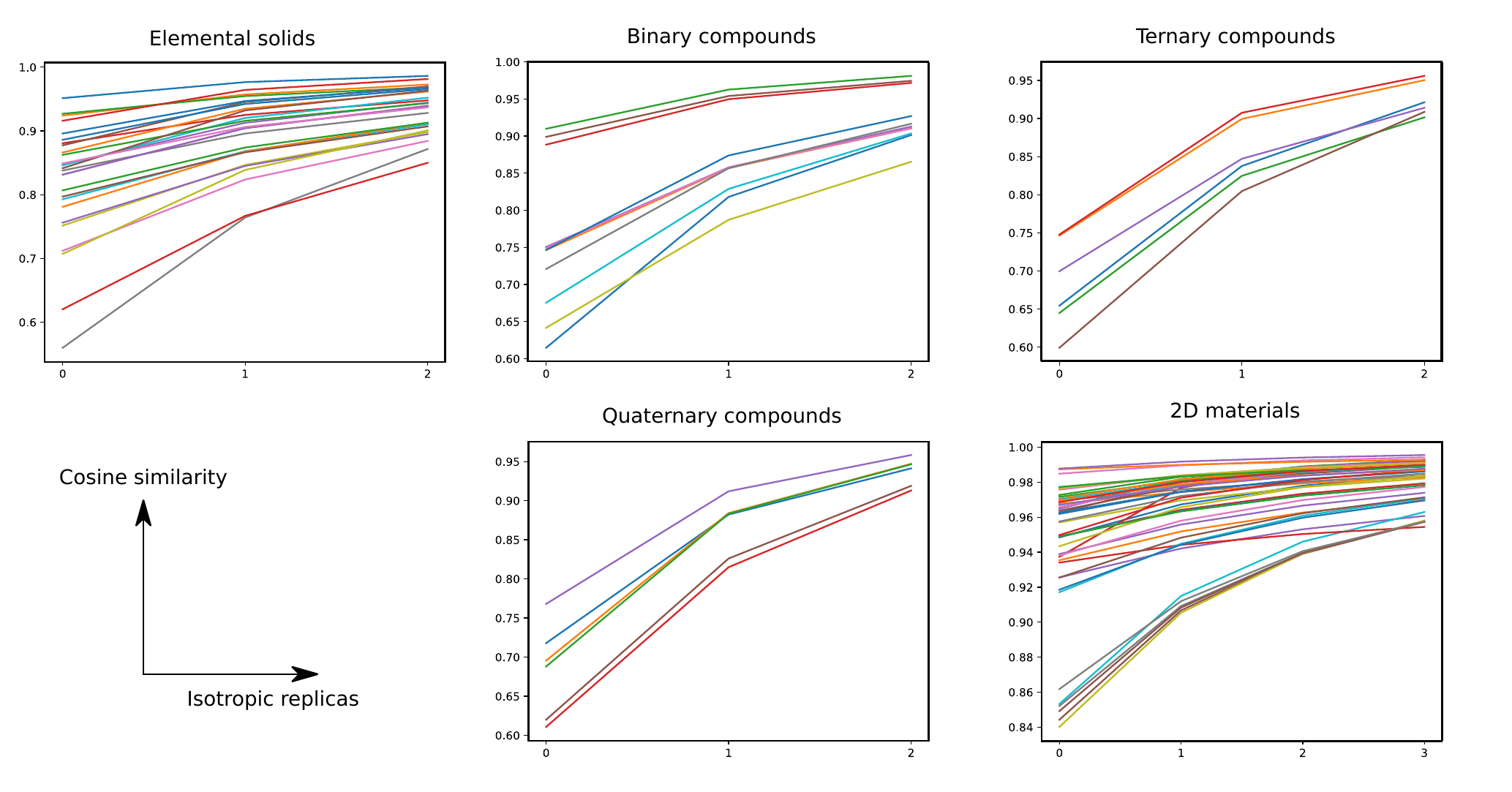}
\caption{Cosine similarity plots for elemental, binary, ternary, and quaternary compounds as well as 2D materials (for SOAP settings $R_\text{C} = 4.0 \cdot d_{\text{NN}}, \sigma = 0.1 \cdot d_{\text{NN}}, $ 
and $\text{exsf} = 1.0$ corresponding to the center of the parameter range used in the training set). Each line corresponds to a particular prototype. 
The x-axis corresponds to three different (non-periodic) supercells, where supercell ``0'' stands for 
the smallest isotropic supercell (for instance $4\times4\times4$ repetitions) for which at least 100 atoms are obtained. Supercells ``1'' and ``2'' correspond to the next two bigger isotropic replicas (e.g., $5\times5\times5$ and 
$6\times6\times6$). The y-axis corresponds to the cosine similarity of the respective supercell to the periodic structure, i.e., the case of infinite replicas. 
One can see a continuous increase of similarity with larger supercell size, where for the largest supercell, the similarity is greater than 0.8 for all prototypes. 
Thus, it is to be expected that systems sizes larger than those included in the training
set can be correctly classified by ARISE. For smaller systems, however, generalization ability will depend on the prototype. Practically, one can  include smaller supercells in the training set, which is not 
 a major problem due to fast convergence time. }
\label{fig:supercells_cosine_sim_to_pbc_True}
\end{figure*}

\textbf{Dataset creation.}\ To compute the training set (39$\,$204 data points in total), we include periodic and non-periodic systems. 
For the former, no supercells are necessary (as SOAP
is supercell-invariant for periodic structures). For the latter, a given structure (or rather its unit cell as obtained from the respective database) is isotropically replicated until at least 100 atoms are contained in the structure. Then this 
supercell structure and the next two larger isotropic replicas are included. 
With this choice of system sizes, we focus on slab- and bulk-like systems.
Note that the network may not generalize to non-periodic structures outside the chosen supercell range.
Practically, if the need to classify much smaller or larger supercells 
arises, one can include additional replicas to the training set and retrain the 
model (while for larger supercells it is expected that the network will generalize, see also Supplementary Fig. \ref{fig:supercells_cosine_sim_to_pbc_True}). Retraining is computationally easy due to fast convergence time. 
Note that for 2D structures, only in-plane replicas are considered. 

Elemental solids and binary compounds are selected from the AFLOW library of crystallographic prototypes\cite{mehl2017aflow}. 
Ternary, quaternary, and 2D materials are taken from the computational materials repository (CMR)\cite{landis2012computational}.

Nanotubes are created using the atomic simulation environment (ASE)\cite{larsen2017atomic} where the chiral numbers (n,m) provide the class labels. 
We filter out chiral indices (n, m)  (with the integer values n,m taking values in $[0,10]$) for which the diameter is in the range $[4\,\mathrm{\AA},6\,\mathrm{\AA}]$ (and skip the cases 
where $n=m=0$, $n<m$). 
Then, we increase the length of each nanotube until at least 100 atoms are contained. No additional lengths are included as it was checked that there is no major change in the SOAP descriptor (via calculating 
the cosine similarity between descriptors representing nanotubes of different length). For more complex nanotubes (for instance, multi-walled systems), this may change.  

For the cutoff $R_{\text{C}}$, we select the range $[3.0 \cdot d_{\text{NN}} , 5.0 \cdot d_{\text{NN}}]$ in steps of $0.2 \cdot d_{\text{NN}}$ and for 
$\sigma$ the values $[0.08 \cdot d_{\text{NN}}, 0.1 \cdot d_{\text{NN}}, 0.12 \cdot d_{\text{NN}}]$. 
We calculate the SOAP descriptor using the quippy  package (\url{https://libatoms.github.io/QUIP}), where 
we choose $n_{\text{max}}=9$ and  $l_{\text{max}}=6$ as limits for 
the basis set expansion, resulting in an averaged SOAP vector of length 316. 
Furthermore, we increase the dataset by varying the so-called extrinsic scaling factor: For a given prototype, the value of $d_{\text{NN}}$ will deviate from the pristine value in presence of defects. Thus, to inform the network that the computation of  $d_{\text{NN}}$ is erroneous, we scale 
each pristine prototype not only by $1.0 \cdot d_{\text{NN}}$ but also $0.95\cdot d_{\text{NN}}$ and $1.05\cdot d_{\text{NN}}$. We term the factors 0.95, 1.0, 1.05 extrinsic scaling factors. 
One may also see this procedure as a way to increase the training set.

To create defective structures, we explained in the main text (cf. Table \ref{table:accuracy-comparison-single-crystal}) how defects (displacements, missing atoms) are introduced. 
Note that we use the term missing atoms and not vacancies since the percentages of removed atoms we consider are well beyond 
regimes found in real materials. Also note that displacements as high as 4\% of the nearest neighbor distance might already cause a transition to the liquid phase in some solids. 
Still, as noted in the Introduction, experimental and computational data often present levels of distortions which are comparable or even substantially exceed these regimes.
We introduce defects for all pristine prototypes included in the training set  (specifically, for the supercells \textendash\ for both periodic and non-periodic 
boundary conditions, while for nanotubes only non-periodic structures are used).
Since the defects are introduced randomly, we run 10 iterations of defect creation on each prototype.   
Then we calculate SOAP for all of these defective structures for one specific parameter setting ($R_{\text{C}} = 4.0 \cdot d_{\text{NN}}, \sigma = 0.1 \cdot d_{\text{NN}}$, extrinsic scaling factor $=1.0$),  
which corresponds to the center of the respective parameter ranges included in the training set. 
Finally, we obtain 5880 defective structures for each defect ratio. In total, we compute defectives structures for three defect types (missing atoms and  displacements introduced both separately and combined) 
for eight different defect ratios, giving in total 141,120 defective data points.

\textbf{Neural-network architecture and training procedure.}\ 
At prediction time, we need to fix 
$T$, the number of forward-passes being averaged (cf. Supplementary Methods). 
We chose $T=10^3$ for all results except Fig. \ref{fig:STEM}c and Supplementary Fig. \ref{fig:assign_most_sim_proto}, for which we increase  $T$ 
to $10^{5}$  in order to get stable assignments in case of high uncertainty and very low probability candidates (i.e., $<1.0\%$). 
Still, 
the most similar prototypes can be obtained already with $10^3$ iterations. 

Training is performed using Adam optimization\cite{kingma2014adam}. 
The multilayer perceptron is implemented in Keras\cite{chollet2015} using Tensorflow\cite{abadi2016tensorflow} as backend. 
Furthermore we optimize hyperparameters such as the number of layers  
using Bayesian optimization, specifically the 
Tree-structured Parzen estimator (TPE) algorithm 
as provided by the python library hyperopt\cite{10.5555/3042817.3042832}
(cf.  Supplementary Methods 
for more details). 

The initial training set is split (80/20\% training / validation split of pristine structures, performed using scikit-learn, in stratified fashion, using a random state of 42) and the accuracy on the validation set is used 
as the performance metric to be minimized via hyperopt (for 50 iterations). 
Fast convergence (followed by oscillations around high accuracy values) or divergence is typically observed, which is why we train for a fixed number of epochs (300) and 
 save only the model with the best performance on the validation set.
Training is performed on 1 GPU (Tesla Volta V100 32GB) on the Talos machine-learning cluster in collaboration with the Max Planck Computing and Data facility (MPCDF). 
We observe that accuracies around $99\%$ can be reached after few iterations, with individual training runs converging within 20 minutes, depending on model complexity. 

Practically, strong models are obtained via this procedure, while further fine-tuning can be made to reach perfect accuracies. 
First, we restrict to one setting of training parameters (see previous section). From a computational efficiency point of view, this is also  the the preferred choice since one has 
to compute only one descriptor per structure during prediction time. 
 We select $R_\text{C}=4.0\cdot d_{NN}$ and $\sigma=0.1\cdot d_{\text{NN}}$ as well as an extrinsic scaling factor of 1.0. These choices
 are at the center of the  respective parameter ranges. 
While the model with highest validation accuracy (on the whole training set) determined via hyperopt usually gives very strong performance, 
it is not necessarily the best possible one,  especially in terms of generalization ability to defective structures. 
To find the optimal (i.e., most robust) model we select some of the best models (e.g.,  top 15) found via hyperopt  
 and rank them based on their performance on pristine and defective structures (again for one setting of $R_\text{C}, \sigma$). 
In particular, we restrict to defective points with either $\leq 5\%$ atoms missing or $<1\% $ atomic displacement, which comprises 35$\,$280 data points (six different defect ratios with 5$\,$880 points each). 
The number of pristine data points is 396. 
Using this strategy, we can identify a model with 100\% accuracy on pristine and defective structures, which is reported in the last line of Table \ref{table:accuracy-comparison-single-crystal}.  
The accuracy on the whole training set comprising 39$\,$204 data points is 99.66\%.

We also investigate the performance on higher defect ratios beyond physically reasonable perturbations, since this is typically encountered 
in atom-probe experiments. 
In particular, we investigate three defect types (missing atoms, displacements, and both of them) comprising 105$\,$840 data points. 
The results for missing atoms ($>5\%$) and displacements ($>0.6\%$) can be found in Table \ref{table:accuracy-comparison-single-crystal} and Supplementary Table \ref{table:suppl_high_defects}. 
Classification accuracies 
on structures with both missing atoms and displacements are specified in Supplementary Table \ref{table:suppl_vac_and_displ}. 
Note that training and model selection only on pristine structures can yield robust models, especially 
if the number of classes is reduced. For instance, training only on binary systems 
using a pristine set of 4$\,$356 data points (full SOAP parameter range) 
gives perfect accuracy on both the full training set and 3$\,$960 defective structures (displacements $\leq 0.06\% $ and $\leq5\%$ missing atoms \textendash\ for the 
setting $R_\text{C}=4.0\cdot d_{\text{NN}}, \sigma = 0.1\cdot d_{\text{NN}}$, extrinsic scaling factor 1.0). Note that in general,  
 if fewer classes are considered (e.g., $\sim$ 20), the training time  can be significantly reduced  (e.g., to a few minutes).

\textbf{Naive Bayes}
We employ the implementation provided by scikit-learn (\url{https://scikit-learn.org/stable/modules/naive_bayes.html}), where two assumptions for the likelihood $P(x_i|y)$
of the features $x_i$ given the labels $y$ are tested: A Gaussian distribution (Gaussian Naive Bayes, short GNB) and a multivariate Bernoulli distribution (Bernoulli Naive Bayes, short BNB).
We observe that the BNB model yields improved results compared to GNB, while both being significantly less accurate than ARISE.

\begin{figure*}[!htb]
\centering
\includegraphics[width=0.9\textwidth]{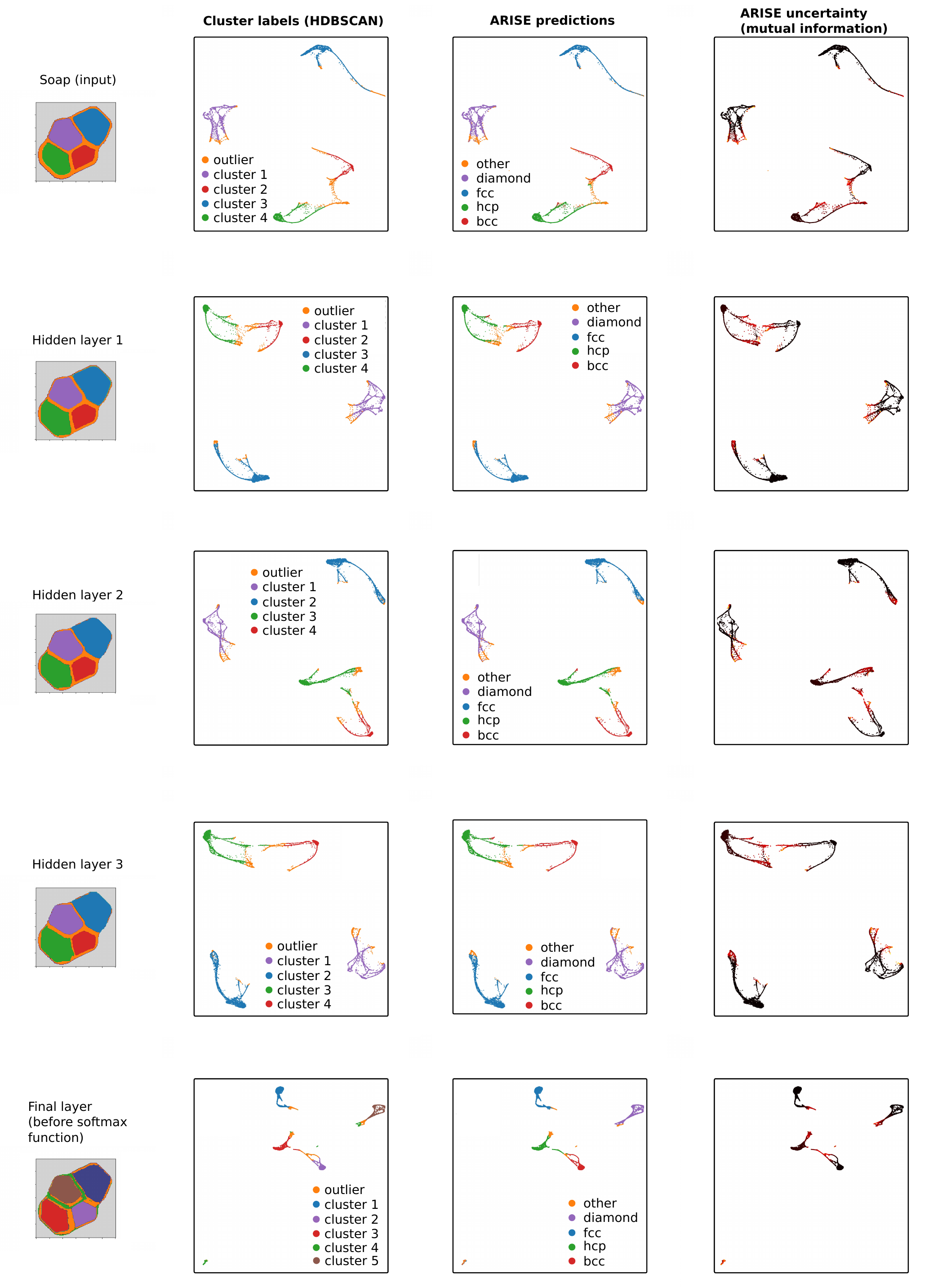}
\caption{Unsupervised analysis analogous to Figure \ref{fig:synthetic_polycrystal}d-g, for all layers (before the ReLU or rather the softmax function) with a minimum distance of 0.1 and a number of neighbors of 500.}
\label{fig:four_grains_umap_hdbscan_full}
\end{figure*}

\textbf{Unsupervised learning: clustering and dimensionality reduction.}\ HDBSCAN\cite{mcinnes2017accelerated, McInnes2017} is a density-based, hierarchical clustering algorithm 
(see also the online documentation \url{https://hdbscan.readthedocs.io/en/latest/}). 
The final (so-called flat) clustering is derived from a hierarchy of clusters. The most influential parameter  is the minimum cluster size that determines the minimum number of data points a cluster has to contain \textendash\ 
otherwise it will be considered an outlier, i.e., not being part of any cluster. 
Practically, one can test a range of values for the minimum cluster size, in particular very small, intermediate and large ones \textendash\ for instance  
for the results on the synthetic polycrystal in Fig. \ref{fig:synthetic_polycrystal}a, we test the  
 values  $\{25, 50, 100, 250, 500, 1\,000\}$. 
 In line with intuition, the number of clusters grows (shrinks) for smaller (larger) values of minimum cluster size.  
 A coherent picture with 4 clusters and clear boundaries (as indicated by the outliers) arises for minimum cluster size values of around 500, for which we report the results in Fig. \ref{fig:synthetic_polycrystal}d-g 
 and Supplementary Fig. \ref{fig:four_grains_umap_hdbscan_full}. 
 Moreover, we test the influence of the so-called minimum distance parameter in Supplementary Fig. \ref{fig:supp_hdbscan_pos_gb}, where for Fig. \ref{fig:synthetic_polycrystal}e-g, we choose a minimum distance parameter of 0.9.   
 
For the nanoparticle data discussed in Fig. \ref{fig:AET}c, we observe that most of the points are considered outliers since the data contains substantially more distortions.  
To address this, we 
use the soft clustering feature of HDBSCAN, which allows to calculate a vector for each data point whose $i$-th component quantifies 
the probability that the given data point is member of cluster $\textit{i}$.
Then, we 
can infer a cluster assignment for points that would normally be considered outliers, by selecting for each point the cluster whose membership probability is maximal (while considering a point an outlier if all probabilities are below a certain threshold for which we choose 10\,\%).
For the minimum cluster size, we find that for values below 10 the number of clusters quickly grows while shrinking for larger values. We report the results for a minimum cluster 
size of 10 and a minimum distance parameter of 0.1 in Fig. \ref{fig:AET}c.

To visualize the clustering results, we use the manifold-learning technique UMAP\cite{mcinnes2018umap} (see also the online documentation \url{https://umap-learn.readthedocs.io/en/latest/}).
This method uses techniques from Riemannian geometry and algebraic topology to capture both the global and local structure of a manifold that underlies a given dataset.
One of the most important parameters 
is the number of neighbors that will be considered to construct a topological representation of the data, where a small value takes only  the local structure into account, while a large value considers the global relations 
between data points. We choose values of 500 for Fig. \ref{fig:synthetic_polycrystal}e-g and 50 for \ref{fig:AET}c, above 
which the 2D embeddings do not change significantly.

\textbf{Synthetic polycrystal generation} The structure in Fig. \ref{fig:synthetic_polycrystal}a is generated via the open-source software Atomsk\cite{hirel2015atomsk}.

\textbf{Strided pattern matching parameters.}\ Two parameters are most important for strided pattern matching analysis:
Firstly, the stride defines the resolution and may be chosen arbitrarily small or large  to increase or decrease the visualization of structural features.
Note that the sliding allows us to discover smooth transitions, while the smoothness is determined by the step size. This way, boundary effects between neighbored local regions 
are reduced compared to the case of slightly or non-overlapping boxes (e.g., in the simple voxelization case). 
In particular, a small stride (e.g., 1\,\AA) mitigates boundary effects due to the discretization, which otherwise can influence  the final classification and uncertainty maps. 
SPM is trivially parallel by construction, thus allowing the time-efficient characterization of large systems. 
Clearly, in a naive implementation, this procedure scales cubically with stride size. Practically,  one may choose a large stride (in particular if
the structure size would exceed computing capabilities) to obtain 
low-resolution classification maps, which may suffice to identify regions of interest. 
Then, one may zoom into these areas and increase the stride to obtain high resolution classification maps revealing more intricate features. 
Secondly, the box size determines the locality, i.e., the amount of structure that is averaged to infer the crystallographic prototype being most similar to a given local region. If this 
parameter is chosen too large, possibly interesting local features may be smoothed out. 
We recommend to use box sizes larger than 10-12$\text{\AA}$, as in these cases, the number of contained atoms is typically within the range of the supercells the network is trained on (i.e., at least 100 atoms). 
The generalization ability to smaller structures depends on the prototype (cf. Supplementary Fig. \ref{fig:supercells_cosine_sim_to_pbc_True}), and in general, if a smaller box size is 
desired while using our model, the practical solution is to add smaller supercells in the training set and retrain the network. 
Note that the shape of the local regions may be chosen to be different from boxes, 
e.g., spheres or any other shape that fits the application at hand. Moreover, we chose the grid in which the structure is strided to be cubic, while 
other discretizations are possible. 
Note that a one-dimensional striding can be applied to rod-like systems such as carbon nanotubes.

In this work, we choose the following SPM parameters: For the slab analysis in Fig. \ref{fig:synthetic_polycrystal}a, we choose a 
$1\, \text{\AA}$ stride and a box size equal to the slab thickness ($16\,\text{\AA}$). 
For the superalloy model system  we choose the same box size but reduce the stride to $3\, \text{\AA}$, since this system is much larger and we want to demonstrate that for these systems, smaller strides 
still yield reasonable results. 
 For the grain-boundary structure in Fig. \ref{fig:synthetic_polycrystal}k, a stride of $2\, \text{\AA}$ and a box size of $10\, \text{\AA}$ suffice to characterize the system. 
For the 2D STEM analysis (cf. Fig. \ref{fig:STEM}g), we choose a stride of 4 (in units of pixels since atoms are reconstructed from images, while for typical graphene bond lengths 
of 1.42\,\AA~ the relation 1\,\AA~ $\approx$ 8.5 can be inferred). Moreover,  we select a box size of 100 pixels ($\approx 12\,\text{\AA}$).
 For the quasicrystalline structure in Fig. \ref{fig:STEM}h,i, which has been cropped from the original reference\cite{li2016review} and rescaled to a $1000\times1000$ pixel image (using standard settings in the GIMP Image editor), a box size of 100 pixels and stride of 10 pixels suffice to detect the MnS$_2$ prototype as dominant pattern. 
For the nanoparticle analysis, we choose a stride of $1\, \text{\AA}$ and box size of $12\,\text{\AA}$ for all of Fig. \ref{fig:AET}, except the clustering analysis, for which 
we reduce the stride to $2\text{\AA}$, to avoid an overcrowded 2D map. 
The box size of $16\,\text{\AA}$ (which allowed 
 to distinguish chemically disordered fcc from ordered L1$_2$, cf. Fig. \ref{fig:synthetic_polycrystal}h-j) yields comparable results (see Supplementary Fig. \ref{fig:supp_nanoparticle_2017_uncertainty}), while finding less L1$_0$ 
 symmetry and more fcc since a larger amount of structure is averaged.
 Due to L1$_0$ showing special magnetic properties, we are interested in having a larger pool of candidate regions, which is why we choose a box size of $12\,\text{\AA}$ (corresponding 
 to the smallest value such that the average number of atoms in each box is greater than  100).

\textbf{Atomic electron tomography.}\

  ARISE's predictions are reliable since all the symmetries that typically occur in FePt nanoparticles are included in the training set \textendash\ except  
 the disordered phase for which it has been demonstrated in the analysis of the Ni-based superalloy model system that ARISE is sensitive to chemical ordering. 
 Moreover, a supplementing study reveals that ARISE can analyze structural transformations, in particular similar to the ones taking place in nanoparticles
 (cf. \nameref{section:supp_note_2} and Supplementary Fig. \ref{fig:Bain_path_resuts}, 
 where the so-called Bain path is investigated).
 
  Due to diffusion, the shape of the three nanoparticles (cf. Fig. \ref{fig:AET}c)  and thus the number of atoms is changing. 
 Rough alignment of the nanoparticles was checked using point set registration:
Specifically, we employed the coherent point drift algorithm\cite{myronenko2010point} as implemented in the python package pycpd  (\url{https://github.com/siavashk/pycpd}). We extracted only the core of the nanoparticle, which is reported to remain similar during the annealing procedure\cite{zhou2020atomic}. 
After applying the algorithm, the remaining mismatch is negligible (3-10$^\circ$ for all three Euler angles).

\subsection*{Data availability}

The training and test data, trained neural-network model, as well as all relevant geometry files and datasets that are 
generated in this study 
have been deposited at Zenodo under accession code \url{https://doi.org/10.5281/zenodo.5526927}. 
The geometry file of the so-called Pearl structure analyzed in Fig. \ref{fig:synthetic_polycrystal}k-m is 
available in Edmond (the Open Access Data Repository of the Max Planck Society) under accession code \url{https://edmond.mpdl.mpg.de/imeji/collection/zV4i2cu2bIAI8B}.
The experimental HAADF image datasets 
and trained neural-network models that are employed in this study for reconstructing atomic positions are available under accession codes \url{https://github.com/pycroscopy/AICrystallographer/tree/master/AtomNet} 
and \url{https://github.com/pycroscopy/AICrystallographer/tree/master/DefectNet}.
 The HRTEM data used in this study (Fig. \ref{fig:STEM}h) 
 has been adapted (see Methods) from the original publication\cite{li2016review}, where it is published under a Creative Commons Attribution 4.0 International License.
The AET data used in this study is available in the Materials Data Bank (MDB) under accession code \url{https://www.materialsdatabank.org/}.

\subsection*{Code availability}

A Python code library ai4materials containing all the code used in this work is available at \url{https://github.com/angeloziletti/ai4materials}. In more detail, ai4materials provides tools to perform complex analysis of materials science data using machine learning techniques. 
Furthermore, functions for pre-processing, saving, and loading of materials science data are provided, with the goal to ease traceability, reproducibility, and prototyping of new models. 
An online tutorial to reproduce the main results presented in this work can be found in the NOMAD Analytics-Toolkit at \url{https://analytics-toolkit.nomad-coe.eu/tutorial-ARISE}.


\noindent

\section*{Acknowledgements}
We acknowledge funding from BiGmax, the Max Planck Society's Research Network on Big-Data-Driven Materials Science. 
L.\ M.\ G.\ acknowledges funding from the European Union's Horizon 2020 research and innovation program, under grant agreements No. 951786 (NOMAD-CoE) and No. 740233 (TEC1p).
Furthermore, the authors acknowledge the Max Planck Computing and Data Facility (MPCDF) for 
computational resources and support, which enabled neural-network training    
on 1 GPU (Tesla Volta V100 32GB) on the Talos machine learning cluster. 
The authors thank Matthias Scheffler for initiating this research direction and providing comments to the manuscript. 

\section*{Author contributions}
All authors designed the project. A.Z. and A.L. wrote the code, A.L. performed all the calculations. A.Z. and L.M.G. supervised the project. All authors wrote and reviewed the manuscript.

\section*{Competing interests}
The authors declare no competing interests.

\clearpage

\beginsupplement

\onecolumn
Supplementary Information for 
 \begin{center}
   \Large{\textbf{Robust recognition and exploratory analysis of crystal structures via Bayesian deep learning}}
 \end{center}

\tableofcontents

\clearpage

\twocolumn

\section{Supplementary Methods}

\textbf{Isotropic scaling.} To reduce the dependency on lattice parameters, we isotropically scale each prototype  according to its nearest neighbor distance $d_{\text{NN}}$. 
This way, one degree 
of freedom is eliminated, implying that all cubic systems 
are equivalent and thus are correctly classified by construction. 
To compute $d_{\text{NN}}$, we calculate 
in a first step the histogram of all nearest neighbor distances. 
Since the area of spherical shells grows with the squared radius, we divide the histogram by the squared radial distance. 
Then, we use the center of the maximally populated bin as the nearest neighbor distance $d_{\text{NN}}$. 
Dividing the atomic position by $d_{\text{NN}}$ yields the final isotropically scaled structure, which is used for calculating the SOAP descriptor. 
Alternatively, one may use the mean of the nearest neighbors as $d_{\text{NN}}$, which, however, is more prone to defects. 
In case of multiple chemical species, we consider all possible substructures as formed by the constituting species to calculate the SOAP descriptor (see next paragraph). For each of the substructures, we 
compute $d_{\text{NN}}$, while we determine the histogram of neighbor distances only from distances between atoms whose chemical species coincide with those of the substructure. 
For instance, given the substructure ($\alpha$, $\beta$), i.e., the atomic arrangement of atoms with species $\beta$ as seen from the perspective of atoms with species $\alpha$, we consider only $\alpha$-atoms and 
determine all distances to $\beta$-atoms.


\textbf{SOAP descriptor.} As discussed in the main text, encoding of  physical requirements we know to be true is crucial for machine-learning application. 
For instance, in crystal classification, 
two atomic structures that differ only by a rotation must have the same classification label. 
This is not guaranteed if real space atomic coordinates 
are used as descriptor (cf. Fig. \ref{fig:single_and_polyc_class_steps}a).
As an attempt to fix this, one might include a discrete subset of orientations in the training set, hoping that the model will generalize to unseen rotations. However, there is no theoretically guarantee that the model will learn the rotational symmetry, and if it does not, it will fail to generalize and return different predictions for symmetrically equivalent structures.
In contrast, when a rotationally invariant descriptor is employed, only one crystal orientation needs to be included in the training set and the model will generalize to all rotations by construction. 
This reasoning readily applies to other physics requirements such as translational, or permutation invariance (for atoms with same chemical species).

In the following, we provide details on adapting the standard SOAP descriptor such that  its number of components is independent on the number of atoms and chemical species. 

Starting with the simple case of one chemical species, we consider a local atomic environment $\mathscr{X}$, being defined  by a cutoff region (with radius $R_\text{C}$) around a central atom, located at the origin of the reference frame. 
Each atom within this area is represented by a Gaussian function centered at the atomic position $\mathbf{r}_i$ and with width $\sigma$. 
Then, the local atomic density function of $\mathscr{X}$ can be written as\cite{bartok2013representing} 
\begin{equation}
 \rho_\mathscr{X}(\mathbf{r})=\sum_{i\in \mathscr{X}} \exp{\left(-\frac{(\mathbf{r}-\mathbf{r}_i)^2}{2\sigma^2}\right)} =\sum_{blm}c_{blm}u_b(r) Y_{lm}(\hat{\mathbf{r}}), \label{equation:local_atomic_density}
\end{equation} 
where in the second step, an expansion in terms of spherical harmonics $Y_{lm}(\hat{\mathbf{r}})$ and a set of radial basis functions $\{ u_b(r)\}$ is performed.  
One can show that the rotationally invariant power spectrum is given by\cite{bartok2013representing} 
\begin{equation}
 p(\mathscr{X})_{b_1b_2l}=\pi\sqrt{\frac{8}{2l+1}}\sum_{m}(c_{b_1 lm})^{\dagger} c_{b_2 lm}. \label{equation:SOAP_power_spectrum}
\end{equation}
These coefficients can be arranged in a normalized (SOAP) vector $\hat{\mathbf{p}}\ (\mathscr{X})$, describing the local atomic environment $\mathscr{X}$. 
In total, we obtain as many SOAP vectors as atoms in the structure, 
which one can average to obtain a materials descriptor independent of the number atoms $N_{\text{at}}$. 
Another possibility (the standard setting in the software we use) is to average the coefficients $c_{blm}$ first and then compute Eq. \ref{equation:SOAP_power_spectrum} from this\cite{mavracic2018similarity}. 
The cutoff radius $R_\text{C}$ and  $\sigma$ (cf. Eq. \ref{equation:local_atomic_density}) 
are hyperparameters, i.e., supervised learning cannot be used directly to assign values to these parameters, while their specific choice will affect the results. Typically, one would employ cross-validation while here, we take a different route: First, we 
assess the similarity between SOAP descriptors using the cosine similarity to 
identify parameter ranges that provide sufficient contrast between the prototypes. Using this experimental approach, we find that values near $\sigma = 0.1 \cdot d_{\text{NN}}$ 
and $R_{\text{C}} =  4.0 \cdot d_{\text{NN}}$ yield good results.  Then we augment our dataset with SOAP descriptors calculated for different 
parameter settings. 

The extension to several chemical species is achieved by considering all possible substructures as formed by the constituting atoms: 
Considering NaCl, we first inspect the lattice of Cl atoms as seen by the Na atoms, 
which we denote by $(\text{Na}, \text{Cl})$; 
this means that Na atoms are considered as central atoms in the 
construction of the local atomic environment while only Cl atoms are considered as neighbors.  
A similar construction is made for the remaining substructures $(\text{Na}, \text{Na})$, $(\text{Cl}, \text{Na})$, and $(\text{Cl}, \text{Cl})$, which may be 
 quite similar depending on the atomic structure. 
For each substructure, we compute the SOAP vectors via Eq. \ref{equation:SOAP_power_spectrum}, obtaining a collection of SOAP vectors. Averaging these gives us four (in case of NaCl) averaged SOAP vectors. 
Averaging the latter again, yields a materials representation being independent on the number of atoms and chemical species.

Formally, given a structure with $S$ species $\alpha_1, ..., \alpha_S$, we consider all substructures formed by pairs of species $(\alpha_i, \alpha_j), j=1,...,S$, resulting in $S^2$ averaged SOAP
vectors $<\hat{\mathbf{p}}_{\alpha_i\alpha_j}>_{N_{\text{at},\alpha_i}}$, where the bracket represents the average over number of atoms $N_{\text{at}}$  of species $\alpha_i$. 
These vectors are  averaged over, yielding the final 
vectorial descriptor  $<<\hat{\mathbf{p}}_{\alpha_i\alpha_j}>>_{\alpha_i\alpha_j}$.

Note that this construction of SOAP deviates from the previously reported way of treating multiple chemical species in the following way: 
Usually, for each atom, one  constructs  
the following power spectra\cite{de2016comparing} 
\begin{equation}
p(\mathscr{X})_{b_1b_2l}^{\alpha\beta}=\pi\sqrt{\frac{8}{2l+1}}\sum_{m}(c_{b_1 lm}^{\alpha})^{\dagger} c_{b_2 lm}^\beta, \label{equation:partial_power_spectra}
\end{equation}
where the coefficients originate from basis set expansion as in Eq. \ref{equation:SOAP_power_spectrum}, while the density $\rho$ is constructed separately for each species. For a specific 
$\alpha$ and $\beta$, the coefficients of Eq. \ref{equation:partial_power_spectra}  can be collected into 
vectors $\mathbf{p}_{\alpha\beta}$. In case of $\alpha\neq\beta$, cross-correlations, i.e., products of coefficients from different densities are used to construct the vectors $\mathbf{p}_{\alpha\beta}$, which  
 are missing in our version.


\textbf{Bayesian deep learning.} As discussed in the main text, one can think of Bayesian neural networks as standard neural networks with distributions being placed placed over the model parameters.
This results in probabilistic outputs from which principled uncertainty estimates can be obtained.
The major drawback is that training and obtaining predictions from traditional Bayesian neural networks is generally difficult because 
it requires solving computationally costly high-dimensional integrals.
For classification, expensive calculations are required to determine $p(y=c|x, \text{D}_{\text{train}})$, which is the probability that the classification is assigned to a class $c$, given input $x$ and training data $\text{D}_{\text{train}}$. 
Then, for a specific input 
$x$ (in our case the SOAP descriptor), the most likely class $c$, i.e., the one with largest $p(y=c|x, \text{D}_{\text{train}})$ is the predicted class.

Gal and Ghahramani\cite{gal2016dropout} showed that stochastic regularization techniques   
 such as   
dropout\cite{hinton2012improving, srivastava2014dropout} can be used to calculate 
high-quality uncertainty estimates (alongside predictions)  
at low cost.
In dropout, 
neurons are randomly dropped in each layer before the network is evaluated for a given input. 
Usually, dropout is only used at training time 
with the goal of avoiding overfitting by preventing over-specialization of individual units. 
Keeping regularization also at test time allows to quantify the uncertainty. 
Practically, given a new input, one collects and subsequently aggregates the predictions while using dropout at prediction time. This gives  
 a collection of probabilities being denoted as $p(y=c|x, \omega_t)$, which is the probability of predicting class $c$ given 
the input $x$ at a specific forward-pass $t$, with model parameters $\omega_t$. 
From this collections of probabilities, one can estimate the actual quantity of interest, $p(y=c|x, \text{D}_{\text{train}})$, 
by a simple average\cite{gal2016dropout}: 
\begin{equation}
p(y=c|x, \text{D}_{\text{train}}) \approx \frac{1}{T} \sum_{t=1}^{T} p(y=c|x,\omega_t),
 \label{equation:approximation_class_probability}
\end{equation}
where $T$ is the number of forward-passes (see Methods section ``Neural network architecture and training procedure'' for details on how we choose this parameter). 
While the average can be used to infer the class label $c$, additional statistical information, which reflects the predictive uncertainty, is contained in the collected forward-passes, i.e., 
the probabilities $p(y=c|x,\omega_t)$ which effectively yield a histogram for each class and define, when varying over all possible $c$, a (discrete) probability distribution. 
For instance, mutual information can be used to quantify the uncertainty from the expressions $p(y=c|x,\omega_t)$. 
Specifically, for a given test point $x$, the mutual information between the predictions and the model posterior $p(\omega|\text{D}_{\text{train}})$ (which captures the most probable parameters given the training data) is defined as\cite{houlsby2011bayesian, gal2016dropout} 
\begin{equation}
\begin{aligned}
& \mathbb{I} \left[y, \omega \vert x, \text{D}_{\text{train}}\right] \approx  \\ 
& - \sum_{c} \left( \dfrac{1}{T} \sum_{t} p \left(y=c \vert x, \boldsymbol{\omega}_t \right) \right)\log  \left( \dfrac{1}{T} \sum_{t} p \left(y=c \vert x, \boldsymbol{\omega}_t \right) \right) \\
& + \frac{1}{T} \sum_{c}\sum_{t} p \left(y=c \vert x, \boldsymbol{\omega}_t \right) \log p \left(y=c \vert x, \boldsymbol{\omega}_t \right). 
\label{equation:mutual_information}
\end{aligned}
\end{equation}


\textbf{Hyperparameter optimization.} The Tree-structured Parzen estimator (TPE) algorithm\cite{bergstra2011algorithms, 10.5555/3042817.3042832}  is an example of a Bayesian optimization technique. 
Specifically, one has to define a search space which can comprise a variety of parameters such as the learning rate 
or model size specifics such as the number of layers or neurons. Then, the goal is to minimize a performance metric (in our case, we  maximize the accuracy by minimizing its negative). 
For large search spaces, iterating through each possible combination, i.e., performing a grid search, will get expensive very quickly. Random search is one alternative, while 
Bayesian methods such as  TPE can be more efficient. 
Approaches such as grid or random search assign uniform probability to each hyperparameter choice,  which implies that a long time is spent at settings with low reward.
This becomes particularly troublesome if the performance metric is expensive to calculate. 
In Bayesian methods such as TPE, the objective is replaced by a computationally cheaper surrogate model (for instance, Gaussian process or random forest regressor). New hyperparameters 
are selected iteratively in a Bayesian fashion. Specifically, the selection 
is based on an evaluation function  (typically so-called expected improvement) taking into account the history of hyperparameter selections and thus avoiding corners of the search space with low reward. 

The search space is chosen the following way (alongside the chosen hyperopt commands hp.choice or hp.uniform):  
\begin{itemize}
 \item Number of layers (2, 3, 4, 5), hp.choice
 \item Number of neurons in each layer (256 or 512), hp.choice
 \item Batch size, (64 or 128), hp.choice
 \item Learning rate, range (0.01,0.0001), hp.uniform
 \item Dropout rate, range (0.01, 0.05), hp.uniform
\end{itemize}

\section{Supplementary Notes}

\subsection{Supplementary Note 1 \label{section:supp_note_1}}

In the following, we provide details on the  benchmarking.

For spglib, we only include prototypes from AFLOW. 
The reason for excluding structures from the computational materials repository (CMR) is that we do not always have the correct or meaningful  labels for all structures. For instance, some 2D materials 
are specified as P1 in the database, 
which cannot be used as a correct label. 
Furthermore, for quaternary chalcogenides, the expected symmetries (as specified in the corresponding reference\cite{pandey2018promising}) cannot be reconstructed, which is most likely due to local optimization effects. 
Similar observations were made for the ternary Perovskites. 
More careful choice of precision parameters or additional local optimization may help. 
Thus, to enable a fair comparison, the benchmarking in the main text only reports results on elemental and binary compounds from AFLOW (where we know the true labels), while the performance on both AFLOW and CMR data is shown in Supplementary Tables  
\ref{table:accuracy-comparison-single-crystal-with-stars}, \ref{table:suppl_high_defects}, and \ref{table:suppl_vac_and_displ}.
To avoid the impression that spglib is not applicable to ternary, quaternary, and 2D materials, we still provide the label ``96/108'' behind spglib methods in the benchmarking tables.
Note that non-periodic structures are excluded for benchmarking (again only in the main table), in particular carbon nanotubes, since these systems cannot be treated by spglib.

For the other benchmarking methods, which are common neighbor analysis (CNA, a-CNA), bond angle analysis (BAA), and polyhedral template matching (PTM), 
we use implementations provided in OVITO\cite{stukowski2009visualization}, where for BAA we apply the Ackland jones method. 
As for spglib, only periodic structures were included.  
BAA, CNA, a-CNA all include fcc, bcc, and hcp structures, while PTM contains in addition sc, diamond, hexagonal diamond, graphene, graphitic boron nitride, L10, L12, zinc blende, and wurtzite. 
Each of the frameworks provide one label for each atom, i.e., for a structure with $N$ atoms we obtain $N$ labels. 
To obtain an accuracy score, we compare these $N$ predictions to $N$ true labels, which correspond to the space group associated with the prototype label (e.g., 194 for hcp).  
For CNA, we select the standard cutoff (depending on its value one is able to detect bcc but not  hcp and vice versa). 
Also for BAA (Ackland jones) and a-CNA standard settings are used. For PTM, an RMSD cutoff of 0.1 was used (again default in OVITO). 
Note that PTM can also distinguish different sites of the L12 structure. For simplicity, we did not label the L12 structure by 
sites and take this classification into account, but always assign a true label as soon as an atom was assigned to the L12 class (even if it might 
be not the correct site).

Furthermore, for ARISE periodic and non-periodic structures are included, while for the benchmarking methods only periodic structures are considered. 
While for spglib, translational symmetry is violated by construction, the other methods can in principle be applied to these systems. 
However, when calculating the accuracy for a given non-periodic structure, we have to choose a label for the boundary atoms. 
If we select the same label for these atoms as for the central ones (which have a sufficiently larger number of neighbors), these methods will usually predict the class ``None'' and 
interpreting this as a ``misclassification'' would decrease the total classification accuracy. Therefore, for a fair comparison, we exclude non-periodic structures.
\clearpage

\begin{table*}[]
\begin{tabular}{@{}lrrrrrrrrrrrr@{}}
\hline \hline
& \multicolumn{1}{c}{Pristine} & \multicolumn{1}{l}{} & \multicolumn{5}{c}{Random displacements ($\delta$)}                                                                                     & \multicolumn{1}{l}{} & \multicolumn{4}{c}{Missing atoms ($\eta$)} \\ 
\cmidrule(lr){4-8} \cmidrule(l){10-13} & & & \multicolumn{1}{c}{0.1\%}  & \multicolumn{1}{c}{0.6\%} & \multicolumn{1}{c}{1\%} & \multicolumn{1}{c}{2\%} & \multicolumn{1}{l}{4\%} & \multicolumn{1}{l}{} & \multicolumn{1}{c}{1\%} &  \multicolumn{1}{c}{5\%} & \multicolumn{1}{c}{10\%} & \multicolumn{1}{c}{20\%}\\ 
\cmidrule(r){1-8} \cmidrule(l){9-13} 
Spglib (loose) & 100.00&& 100.00  & 100.00  & 95.26  & 0.20 &  0.00 && 11.23 &  0.00  & 0.00 & 0.00   \\
 & &&   &   &   &  &   &&  &    &  &    \\
Spglib* (loose) & 67.71 &&  67.71 &  67.71 &  65.83 & 14.51 &  0.00 && 15.73 & 0.03   & 0.00  & 0.00  \\
 & &&   &   &   &  &   &&  &    &  &    \\
Spglib (tight) & 100.00 &&  0.00 &  0.00 &  0.00 & 0.00 &  0.00 && 11.23 & 0.00   & 0.00  & 0.00  \\
 & &&   &   &   &  &   &&  &    &  &    \\
Spglib* (tight) & 83.33 &&  0.00 &  0.00 &  0.00 & 0.00 &  0.00 && 17.53 & 0.00   & 0.00  & 0.00  \\ 
 & &&   &   &   &  &   &&  &    &  &    \\
PTM & 100.00 && 100.00 & 100.00 & 100.00 & 100.00 & 100.00 && 88.67 & 51.76 & 25.93 & 6.24 \\
 & &&   &   &   &  &   &&  &    &  &    \\
PTM* & 8.78 && 11.37 & 11.37 & 11.37 & 11.37 & 11.37 && 10.08 & 5.90 & 2.96 & 0.71 \\ 
 & &&   &   &   &  &   &&  &    &  &    \\
CNA & 66.14 && 62.81 & 62.81 & 54.55 & 32.34 & 31.41 && 55.86 & 32.50 & 15.75 & 3.07 \\
 & &&   &   &   &  &   &&  &    &  &    \\
CNA* & 1.44 && 1.62 & 1.62 & 1.40 & 0.83 & 0.81 && 1.44 & 0.84 & 0.41 & 0.08 \\
 & &&   &   &   &  &   &&  &    &  &    \\
a-CNA & 100.0 && 100.0 & 100.0 & 100.0 & 100.0 & 100.0 && 89.25 & 52.81 & 25.92 & 5.37 \\
 & &&   &   &   &  &   &&  &    &  &    \\
a-CNA* & 2.49 && 3.08 & 3.08 & 3.08 & 3.08 & 3.08 && 2.75 & 1.64 & 0.81 & 0.17 \\ 
 & &&   &   &   &  &   &&  &    &  &    \\
BAA & 100.0 && 100.0 & 100.0 & 100.0 & 100.0 & 97.85 && 99.71 & 88.78 & 65.21 & 25.38 \\ 
 & &&   &   &   &  &   &&  &    &  &    \\
BAA* & 2.49 && 3.08 & 3.08 & 3.08 & 3.08 & 3.03 && 3.08 & 2.74 & 2.02 & 0.81 \\
 & &&   &   &   &  &   &&  &    &  &    \\
GNB  &62.63 && 56.50 & 55.94 & 55.56& 54.98  & 52.72  &&  54.51  &52.94 & 52.67 & 52.09 \\
 & &&   &   &   &  &   &&  &    &  &    \\
BNB  &75.76 && 65.56 & 65.19 & 63.61& 61.58  & 56.58  &&  65.49  &64.00 & 62.43 & 60.48 \\
 & &&   &   &   &  &   &&  &    &  &    \\
\textbf{ARISE} (this work)  &100.00 && 100.00 & 100.00 & 100.00& 99.86  & 99.29  &&  100.00  &100.00 & 100.00 & 99.85 \\ 

\hline \hline
\end{tabular}
\caption{Accuracy in identifying the parent class of defective crystal structures. 
Two lines are shown for each of the methods used for benchmarking (spglib, PTM, CNA, a-CNA, BAA): In rows without stars, the accuracy is calculated only for structures for which the respective method was designed for; 
for instance, spglib can be applied to every structure of Fig. \ref{fig:single_and_polyc_class_steps}e except the 12 nanotubes (note that we only include prototypes from AFLOW for spglib, cf. Supplementary Note 1). 
This is also true for the other methods, while additional structures have to be removed for instance for 
CNA, a-CNA, and BAA as they cannot classify simple cubic and diamond structures. 
In starred rows, all 108 classes summarized in Fig. \ref{fig:single_and_polyc_class_steps}e are included, leading to the strong decrease in performance.  
In contrast, the neural network approach proposed here can be applied to all classes,
and thus the whole dataset was used. 
 Moreover, we compare ARISE to a standard Bayesian approach: Naive Bayes (NB). We consider two different variants of NB: Bernoulli NB (BNB) and Gaussian NB (GNB), where the whole dataset was used \textendash\ see the Methods section 
for more details. ARISE is 
overwhelmingly more accurate 
than both NB methods, for both pristine and defective structures.
}
\label{table:accuracy-comparison-single-crystal-with-stars}
\end{table*}

\begin{table*}[]
\centering
\begin{tabular}{@{}lrrrrrrrrrrrr@{}}
\hline \hline
 & \multicolumn{1}{l}{} & \multicolumn{2}{c}{Random displacements ($\delta$)}  & \multicolumn{1}{l}{} & \multicolumn{2}{c}{Missing atoms ($\eta$)} \\ 
\cmidrule(lr){3-4} \cmidrule(l){5-7} &  & \multicolumn{1}{c}{7\%}  & \multicolumn{1}{r}{10\%} &  \multicolumn{1}{l}{} & 
\multicolumn{1}{c}{25\%} &  \multicolumn{1}{c}{30\%}   \\ 
\cmidrule(r){1-4} \cmidrule(l){5-7} 
Spglib (loose) && 0.00  & 0.00     && 0.00 &  0.00  \\
 &&   &      &&  &    \\
Spglib* (loose)  &&  0.00 &  0.00  && 0.00 & 0.00      \\
 &&   &      &&  &    \\
Spglib (tight)  &&  0.00 &  0.00 && 0.00 & 0.00   \\
 &&   &      &&  &    \\
Spglib* (tight)  &&  0.00 &  0.00  && 0.00 & 0.00      \\
 &&   &      &&  &    \\
PTM && 100.00 & 94.34 && 3.33 & 1.72 \\
 &&   &      &&  &    \\
PTM* && 11.37 & 10.71 && 0.38 & 0.19 \\
 &&   &      &&  &    \\
CNA && 31.41 & 24.20 && 1.38 & 0.55 \\
 &&   &      &&  &    \\
CNA* && 0.81 & 0.62 && 0.04 & 0.01 \\
 &&   &      &&  &    \\
a-CNA && 99.99 & 94.55 && 2.60 & 1.03 \\
 &&   &      &&  &    \\
a-CNA* && 3.08 & 2.90 && 0.08 & 0.03 \\
 &&   &      &&  &    \\
BAA && 87.79 & 69.68 && 14.25 & 7.35 \\
 &&   &      &&  &    \\
BAA* && 2.77 & 2.22 && 0.49 & 0.30 \\
 &&   &      &&  &    \\
GNB  && 50.73 & 48.62    &&  51.33  &50.32   \\
 &&   &      &&  &    \\
BNB  && 48.81 & 43.28    &&  59.78  &58.18   \\
 &&   &      &&  &    \\
\textbf{ARISE} (this work)  && 97.82 & 94.56    &&  99.86  &99.76   \\

\hline \hline
\end{tabular}
\caption{Accuracy in identifying the parent class of defective crystal structures for high displacements (percentage $\delta$) and missing atoms (percentage $\eta$). }
\label{table:suppl_high_defects}
\end{table*}

\begin{table*}[]

\begin{tabular}{@{}lrrrrrrrrrrrr@{}}
\hline \hline
 & \multicolumn{1}{l}{} && \multicolumn{8}{c}{Missing atoms and displacements ($\eta$, $\delta$)}   \\ 
\cmidrule(lr){3-9} \cmidrule(l){9-12} &  \multicolumn{1}{c}{(1\%, 0.1\%)}  & \multicolumn{1}{c}{(5\%, 0.6\%)} & \multicolumn{1}{c}{(10\%, 1\%)} & \multicolumn{1}{c}{(15\%,2\%)} & \multicolumn{1}{l}{(20\%,4\%)} 
& \multicolumn{1}{l}{(25\%,7\%)} & \multicolumn{1}{l}{(30\%,10\%)}    \\ 
\cmidrule(r){1-8} 
Spglib (loose) & 11.32  & 0.00  & 0.00  & 0.00 &  0.00 & 0.00 & 0.00    \\
 &   &   &   &  &   &  &     \\
Spglib* (loose)  &  15.76 &  0.00 &  0.00 & 0.00 &    0.00 & 0.00 & 0.00    \\
 &   &   &   &  &   &  &     \\
Spglib (tight)  &  0.00 &  0.00 &  0.00 & 0.00 &  0.00   & 0.00 & 0.00   \\
 &   &   &   &  &   &  &     \\
Spglib* (tight)  &  0.00 &  0.00 &  0.00 & 0.00 &  0.00   & 0.00 & 0.00 \\ 
 &   &   &   &  &   &  &     \\
PTM & 88.68 & 51.78 & 25.60 & 12.75 & 6.41 & 3.19 & 1.46 \\
 &   &   &   &  &   &  &     \\
PTM* & 10.08 & 5.90 & 2.92 & 1.45 & 0.73 & 0.36 & 0.16 \\ 
 &   &   &   &  &   &  &     \\
CNA & 55.77 & 31.95 & 13.83 & 4.41 & 2.03 & 0.79 & 0.19 \\ 
 &   &   &   &  &   &  &     \\
CNA* & 1.44 & 0.82 & 0.36 & 0.11 & 0.05 & 0.02 & 0.00 \\ 
 &   &   &   &  &   &  &     \\
a-CNA & 89.21 & 52.36 & 26.01 & 12.13 & 6.07 & 2.40 & 0.97 \\ 
 &   &   &   &  &   &  &     \\
a-CNA* & 2.75 & 1.62 & 0.81 & 0.38 & 0.19 & 0.08 & 0.03 \\ 
 &   &   &   &  &   &  &     \\
BAA & 99.72 & 88.98 & 65.17 & 42.62 & 25.95 & 15.58 & 6.63 \\ 
 &   &   &   &  &   &  &     \\
BAA* & 3.07 & 2.75 & 2.02 & 1.34 & 0.82 & 0.50 & 0.22 \\
 &   &   &   &  &   &  &     \\
GNB  & 54.92 & 52.86 & 52.11 & 50.70  & 49.92 & 47.94 & 42.65   \\
 &   &   &   &  &   &  &     \\
BNB  & 64.44 & 61.79 & 58.86 & 56.26  & 52.31 & 45.03 & 40.14   \\
 &   &   &   &  &   &  &     \\
\textbf{ARISE} (this work)  & 100.00 & 100.00 & 100.00 & 99.88  & 99.29 & 97.31 & 92.50   \\
\hline \hline
\end{tabular}
\caption{Accuracy in identifying the parent class of defective crystal structures, with both missing atoms (percentage $\eta$) and displacements (percentage $\delta$) introduced at the same time. 
The results show that ARISE is also robust for highly defective structures where displacements and missing atoms are present at the same time. 
}
\label{table:suppl_vac_and_displ}
\end{table*}


\begin{figure*}[!htbp]
  \begin{center}
    \includegraphics[width=\textwidth]{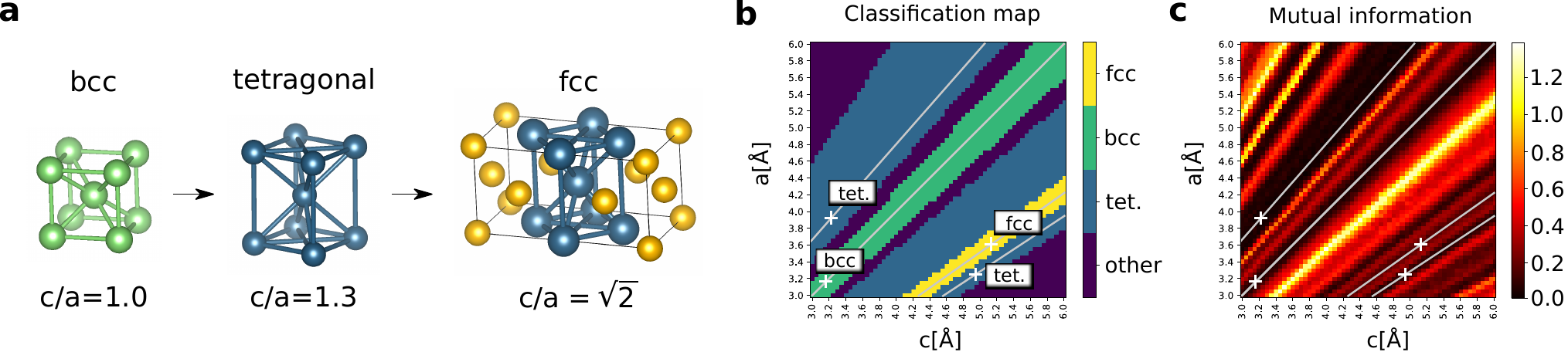}
  \end{center}
  \caption{Application of ARISE for the characterization of the Bain path. \textbf{a} Structures occurring in the Bain path, obtained by varying c/a; increasing c/a from 1.0 (bcc) leads to transitions to tetragonal phases and finally to the fcc structure (c/a=$\sqrt{2}$) .
  \textbf{b} Classification (argmax predictions, left) and uncertainty (mutual information, right) for geometries in the range $c,a \in [3.0\,\text{\AA}, 6.0 \text{\AA}]$.  
  Geometries included in the training set are marked by stars 
  in \textbf{b,c}. As we isotropically scale the structures, 
  geometries with constant c/a are equivalent, which is indicated by solid lines.}
  \label{fig:Bain_path_resuts}
\end{figure*}

\subsection{Supplementary Note 2 \label{section:supp_note_2}}

The Bain transformation path 
describes a structural transitions between bcc and fcc symmetries via intermediate tetragonal phases\cite{bain1924nature} of body-centered \textendash\ or equivalently \textendash\ face-centered
tetragonal symmetry. 
Originally investigated for iron\cite{bain1924nature}, the Bain path is relevant in thermo-mechanical 
processing  \textendash\ a central aspect for steel properties\cite{zhao1995continuous}  \textendash\ as it serves as a model for temperature-induced transitions between fcc $(\gamma)$ and bcc ($\alpha$) iron\cite{grimvall2012lattice}. 
The Bain path is also crucial for understanding properties of epitaxial films\cite{scheffler_bain_path, 2nd_most_prominent_bain_path_paper} or metal nanowires\cite{bain_path_dft_metal_nanowires}.

Practically, the structures constituting a Bain path can be obtained by varying the ratio $c/a$ between lattice parameters $a$ 
and $c$ of a tetragonal structure (cf. Supplementary Fig. \ref{fig:Bain_path_resuts}a); $c/a=1$ corresponds to a cubic structure. 
We generate tetragonal geometries for lattice parameters $a,c$ taking values in $[3.0\,\text{\AA}, 6.0\,\text{\AA}]$ with steps of 0.05\,\AA, resulting in 3721 crystal structures. These structures are then classified with ARISE, 
and the results depicted via 
classification and uncertainty maps in Supplementary Fig. \ref{fig:Bain_path_resuts}b and c, respectively.  Each point in these maps corresponds to a prediction for a specific geometry. 
We include in the training set fcc, bcc, and tetragonal geometries with structural parameters known experimentally; they are shown as stars in Supplementary Fig. \ref{fig:Bain_path_resuts}b. 
Specifically, the lattice parameters $(a, c, c/a)$ are   
$(3.155\text{\AA}, 3.155\text{\AA}, 1.0)$ for the bcc\cite{davey1925lattice} and $(3.615 \text{\AA}, 5.112 \text{\AA}, \sqrt{2})$ for the fcc prototype\cite{straumanis1969lattice}, while two tetragonal structures (being assigned one common label ``tetragonal'') 
are included with  
 $(3.253 \text{\AA}, 4.946, \text{\AA}, 1.521$) 
in case of In\cite{deshpande1969anisotropic}  and  $(3.932 \text{\AA}, 3.238 \text{\AA}, 0.824)$ for $\alpha-\text{Pa}$\cite{zachariasen1959crystal}. 
We isotropically scale every geometry to remove one degree of freedom (see Supplementary Methods section), so that all possible cubic lattices are effectively equivalent; this allows the model to generalize by construction to all cubic lattices regardless of the lattice parameter. The same holds for tetragonal structures (i.e., two degrees of freedom) with constant $c/a$ ratio. 
As visual aid, we mark lines 
of constant c/a in Supplementary Fig. \ref{fig:Bain_path_resuts}b-c starting from the four structures included in the training set.  
Note that any path connecting the constant c/a ratios corresponding to fcc and bcc structures constitutes a \text{Bain path}. 
To obtain a classification label, we select the class with the higher classification 
probability through a so-called argmax operation (i.e., the label $c$ maximizing Eq. \ref{equation:approximation_class_probability}). These predictions are shown in Supplementary Fig. \ref{fig:Bain_path_resuts}. 

The model is able 
to detect the bcc and fcc phases in the expected areas, while all prototypes not being fcc, bcc, or tetragonal are correctly labeled as ``Other''. We point out that only four structures \textendash\ corresponding to points in the plot marked by the four stars \textendash\ are included 
in the training set, while all other 3717 
structures are model (test) predictions. 
We can also observe that the model correctly predict the presence of a a tetragonal phase between fcc (yellow band) and bcc (green band), even though no tetragonal structures from this region are included in the training set. This transition is smooth, only interrupted by small areas for which other, low-symmetry prototypes are assigned, but with high uncertainty, as quantified by the mutual information, cf. Supplementary Fig. \ref{fig:Bain_path_resuts} c. 
We provide the classification probabilities of all assigned prototypes in Supplementary Fig. \ref{fig:suppl_bain_path}. 
In general, increased uncertainty appears at transitions between the assignments of different prototypes. 
We also note that there is a smooth transition for classification probabilities at the transition between prototypes (cf. Supplementary Fig. \ref{fig:suppl_bain_path}). 
These results represent a first indication that the network has learned physically meaningful representations. 
Surprisingly, for large or small $c/a$ ratios, i.e., points far outside the training set, other (low-symmetry) phases appear such as base-centered orthorhombic molecular iodine or face-centered orthorhombic $\gamma-$Pu 
with small uncertainty. 
While it may be desirable  to avoid overconfident predictions far away from the training set, 
the assignments could be actually physically justified given the 
similarities between tetragonal and orthorhombic lattices, the most evident being that all angles in both crystal systems are $90^\circ$. 
We note that the transition to these prototypes is encompassed by regions of high uncertainty also in this case in agreement with physical intuition.


\begin{figure*}[!htb]
\centering
\includegraphics[width=\textwidth]{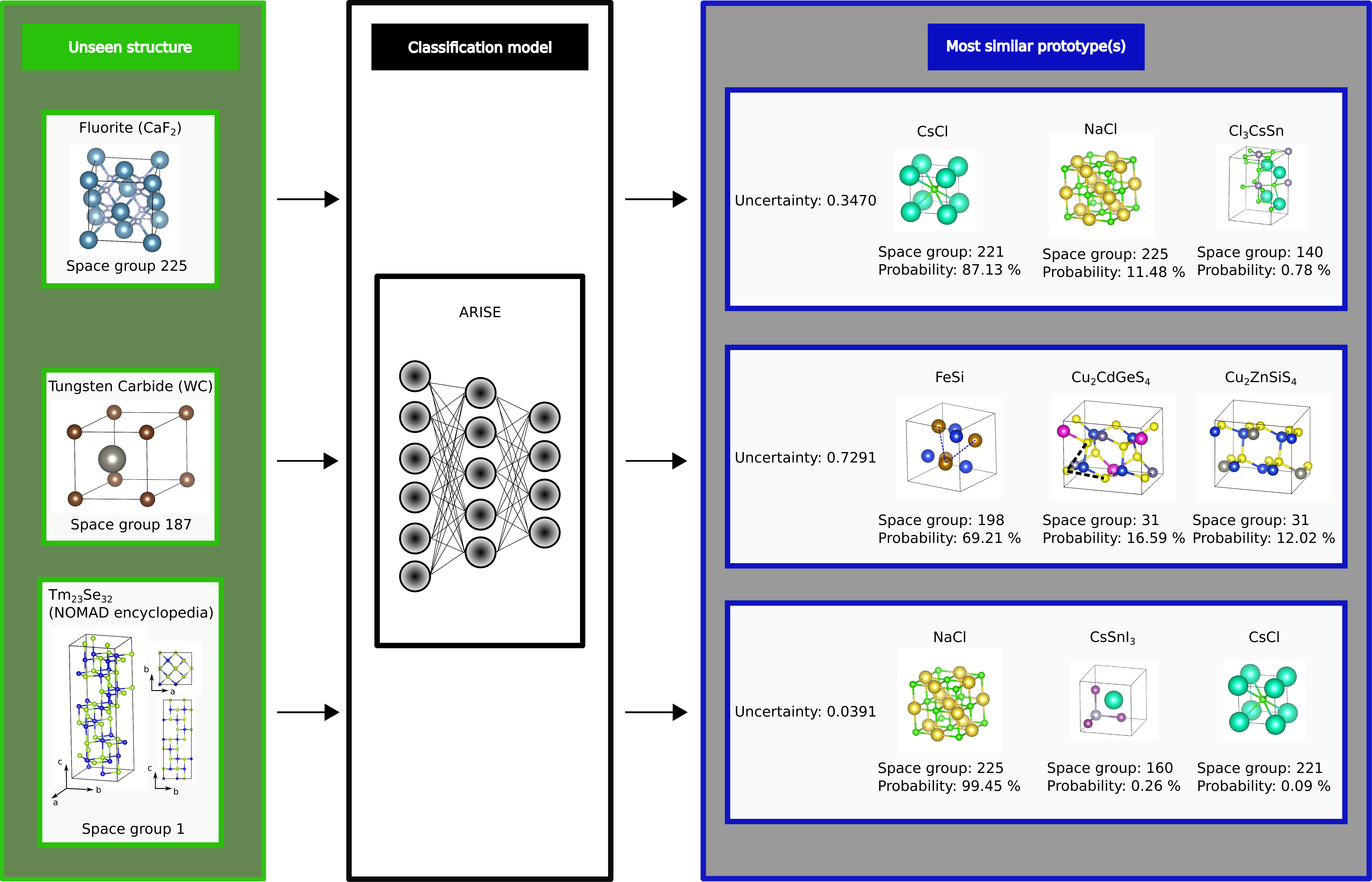}
\caption{
Three examples for assigning the most similar prototype(s) (right panel) to structures for which the corresponding structural class is not contained in the training set of ARISE (left panel). 
For each prediction,  space group 
and  classification probabilities of the top predictions are specified together with an uncertainty estimate (mutual information). 
The space groups are returned via spglib, where we choose the highest symmetry that is found for all combinations of precision parameters $(0.1, 0.01, 0.001, 0.0001) [\text{\AA}]$ and  angle tolerances $(1, 2, 3, 4, 5)[ ^\circ]$. 
}
\label{fig:assign_most_sim_proto}
\end{figure*}

\subsection{Supplementary Note 3 \label{section:supp_note_3}}

In the following, we investigate scenarios in which the model is forced to fail, i.e., we analyze ARISE out-of-sample predictions.

To assess the physical content learned by the network, 
 we investigate its predictions \textendash\ and thus its generalization ability \textendash\ on structures corresponding to prototypes not included in the training.   
This is of particular relevance if one wants to use predictions of ARISE \textendash\ for applications such as screening of large databases, or create low-dimensional maps for a vast collection of materials\cite{isayev2015materials}.

Given an unknown structure, the network needs to decide \textendash\ among the classes it has been trained on \textendash\ which one is the most suitable. It will assign the most similar prototypes and quantify the similarity via classification probabilities, providing a ranking of candidate prototypes. 
The uncertainty in the assignment, as quantified by mutual information, measures the reliability of the prediction. 
Note that the task of assigning the most similar prototype(s) to a given structure among 108 possible classes (and quantifying the similarity) is a very complicated task even for trained materials scientists, in particular in case of complex periodic and possibly defective three-dimensional structures.

We consider three examples (cf. Supplementary Fig. \ref{fig:assign_most_sim_proto} left): 
fluorite and tungsten Carbide (from AFLOW) where the correct labels are known, and one structure from the NOMAD encyclopedia (see last paragraph of this section for details on the provenance), 
for which the assigned space group is 1, i.e., no symmetry can be identified (via spglib). 
In all three cases there is no prototype in the dataset which represent a match for any of these structures. 
This is on purpose: the network will ``fail'' by construction since the correct class is not included in the possible classes the network knows (and needs to choose from). Analyzing how the network fails 
will give us insight on the physical content of the learned model.
This test can also be viewed as discovering ``unexpected similarities'' across materials of different chemical composition and dimensionality. 

Following the pipeline for single-crystal classification 
summarized in Fig. \ref{fig:single_and_polyc_class_steps}, we compute classification probabilities and mutual information, yielding the assignments shown in Supplementary Fig. \ref{fig:assign_most_sim_proto} right.  
To rationalize the predictions shown in Supplementary Fig. \ref{fig:assign_most_sim_proto} from a physical standpoint, we inspect the substructures formed by the chemical species in both original and assigned structures. 
This is motivated by our choice of materials representation as averaged SOAP descriptor of substructures (see Supplementary Methods for more details). 
The two most similar prototypes to fluorite (Ca$\text{F}_2$) are CsCl and NaCl, both consisting of two inter-penetrating  lattices of the same type, two sc lattices for CsCl and two fcc lattices for NaCl. 
Fluorite contains both sc (F atoms) and fcc (Ca atoms) which is likely why CsCl and NaCl are assigned, together with a ternary halide tetragonal perovskite, also containing sc symmetry (via Cs and Sn atoms, respectively). 

For tungsten carbide (WC), W and C form two hexagonal lattices. 
In the unit cell of the most similar prototype, FeSi,  60$^\circ$ angles are formed within the substructures of each species (see dashed lines in the unit cell), thus justifying this classification. 
Furthermore, two quaternary chalcogenides appear as further candidates.  
This similarity \textendash\ hard to assess by eye \textendash\ originates by the presence of angles close to 60$^\circ$ for $S$ atoms (yellow) for both $\text{Cu}_2\text{CdGeS}_4$ and $\text{Cu}_2\text{ZnSiS}_4$ (marked in the figure for $\text{Cu}_2\text{CdGeS}_4$). 
Also note that these two quaternary prototypes, $\text{Cu}_2\text{ZnSiS}_4$ and  $\text{Cu}_2\text{CdGeS}_4$ are a result of substituting Ge and Si with isoelectric elements Zn and Cd, which implies that these structures are expected to be similar. 
This explains why they both appear as candidates for structures being similar to tungsten carbide. 
Finally, for the compound $\text{Tm}_{\text{23}}\text{Se}_{\text{32}}$ from the NOMAD encyclopedia, the model identifies NaCl as the most similar prototype. 
Looking at the structure from different angles, especially from the top (cf. Supplementary Fig. \ref{fig:assign_most_sim_proto}, left part), a similarity to cubic systems 
can be identified. 
The classification method robustness to missing atoms makes the apparent gaps in the side-view negligible, and thus rationalizes the  NaCl assignment. 
Regarding the uncertainty quantification (via mutual information), increased uncertainties appear for fluorite and tungsten carbide, since besides the top prediction with more than $70\%$ classification 
probability, other  prototypes are possible  candidates for the most similar prototype. 
For the NOMAD structure $\text{Tm}_{\text{23}}\text{Se}_{\text{32}}$, the 
network is quite confident, most likely because no other good candidates are presented among the binaries included in the 108 classes dataset. 

These results show that the model \textendash\ even when forced to fail by construction \textendash\ returns (highly non-trivial) physically meaningful predictions. 
This makes ARISE particularly attractive for screening large and structurally diverse databases, in particular assessing structures for which no symmetry label can be obtained with any of the current state-of-the-art methods.

In addition to the analysis in Supplementary Fig. \ref{fig:assign_most_sim_proto}, we present some results for further out-of-sample structures: 
\begin{itemize}
 \item Boron nitride (bulk, graphitic, \url{http://aflowlib.org/CrystalDatabase/AB_hP4_194_c_d.html}) classified as hexagonal graphite (probability 63.32\%), mut.inf. 0.7278
 \item Cementite (\url{http://aflowlib.org/CrystalDatabase/AB3_oP16_62_c_cd.html}) classified as MnP with probability 49.14\%, mut.inf. 0.7176
 \item $\text{CuTi}_3$ (L60 Srukturbericht, \url{http://aflowlib.org/CrystalDatabase/AB3_tP4_123_a_ce.html}) classified as bct $\alpha$-Pa with probability 78.41\%, mut.inf. 0.8539 
 \item Benzene (\url{http://aflowlib.org/CrystalDatabase/AB_oP48_61_3c_3c.html}) classified as nanotube (chiral indices (n,m)=(5,2)) with probability 68.48\%, mut.inf. 0.6249
 \item Rutile (\url{http://aflowlib.org/CrystalDatabase/A2B_tP6_136_f_a.html}) classified as orthorhombic halide perovskite with probability 44.62\%, mut.inf. 0.8733
 \item NbO (\url{http://aflowlib.org/CrystalDatabase/AB_cP6_221_c_d.html}), which is NaCl with 25\% ordered vacancies on both the Na and Cl sites,  classified as NaCl with probability 99.96\%, mut.inf. 0.0027
 \item Moissanite 4H SiC (\url{http://aflowlib.org/CrystalDatabase/AB_hP8_186_ab_ab.html}) classified as wurtzite with probability 99.74\%, mut.inf. 0.0166 
 \item $\text{K}_2\text{PtCl}_6$ (\url{http://aflowlib.org/CrystalDatabase/A6B2C_cF36_225_e_c_a.html}) classified as NaCl with probability 61.4\%, mut.inf. 0.5402 
\end{itemize}

The structure taken from the NOMAD encyclopedia 
has the ID mp-684691 in Materials project, where further details can be found, e.g., on
on   
the experimental origin. 

\begin{figure*}[!htb]
\centering
\includegraphics[width=\textwidth]{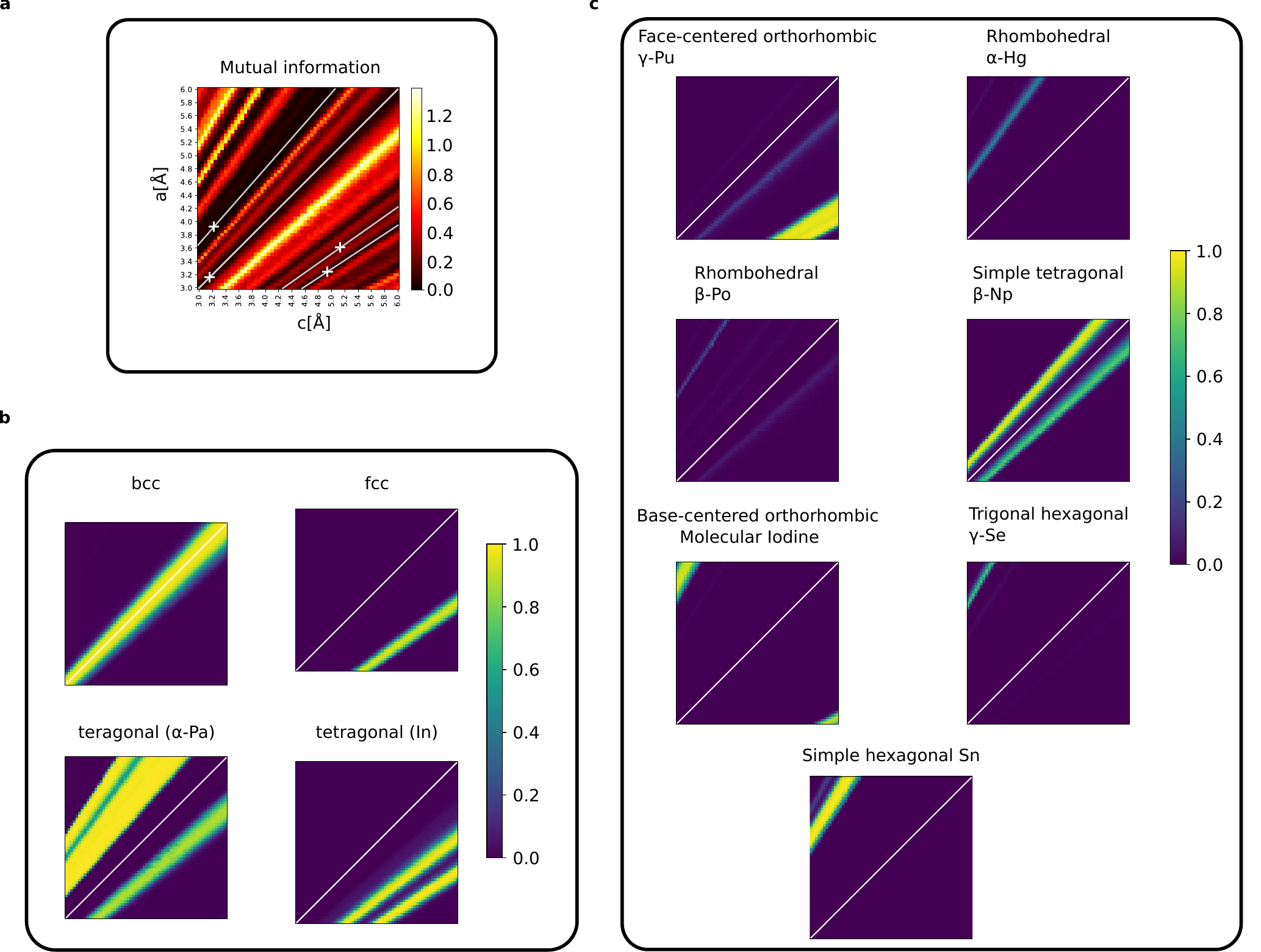}
\caption{Bain path - all prototypes with increased classification probability: \textbf{a} Mutual information plot showing the spots of high and low uncertainty for different geometries. 
\textbf{b} Classification probability maps corresponding to bcc, fcc and two tetragonal phases. 
\textbf{c} Representative selection of other prototypes showing non-zero classification probabilities. }
\label{fig:suppl_bain_path}
\end{figure*}

\begin{figure*}[!htb]
\centering
\includegraphics[width=\textwidth]{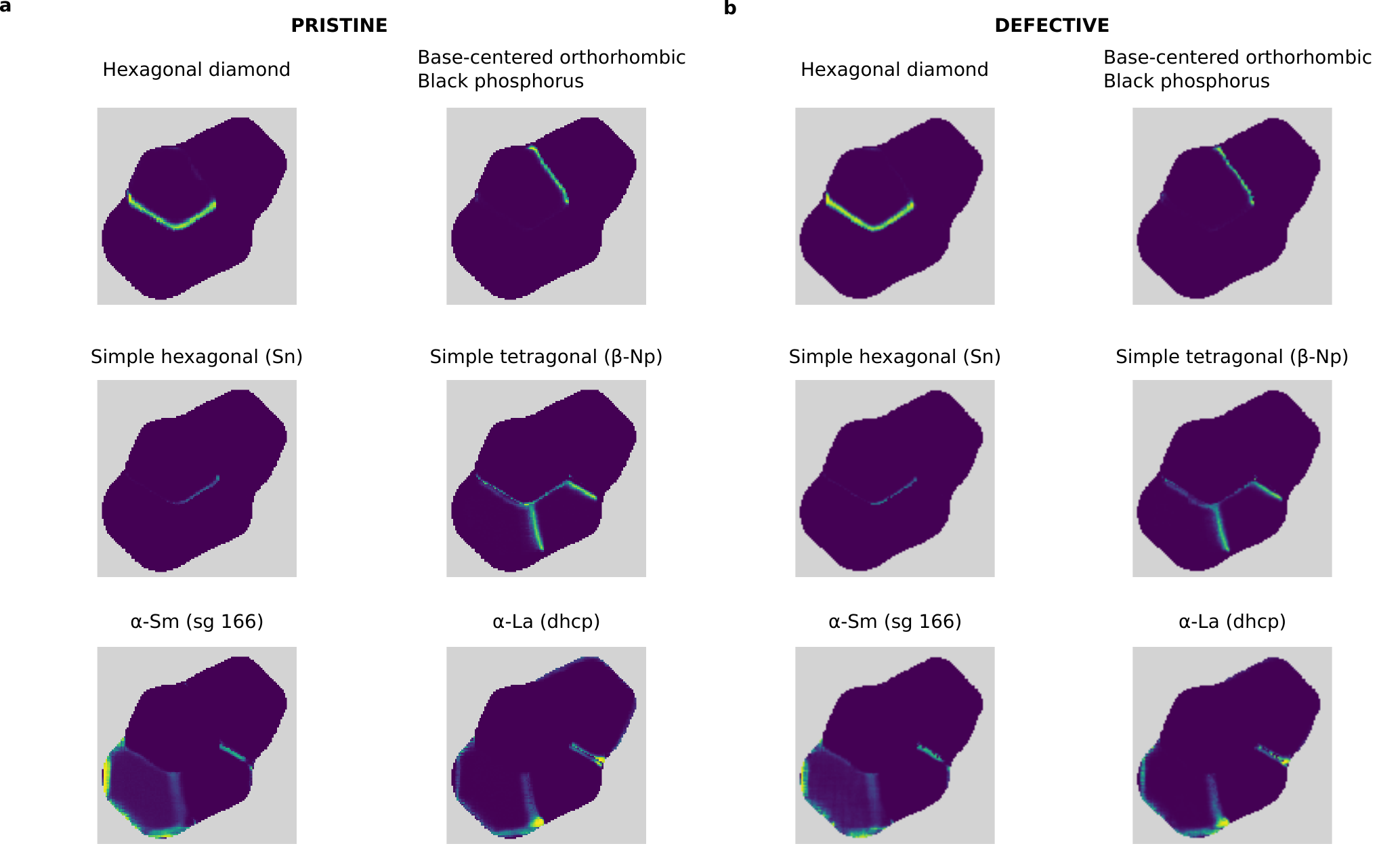}
\caption{Probability maps of the most important prototypes for both pristine (\textbf{a}) and defective (\textbf{b}) version of the mono-species polycrystal in Fig. \ref{fig:synthetic_polycrystal}a. 
}
\label{fig:suppl_four_grains}
\end{figure*}

\begin{figure*}[!htb]
\centering
\includegraphics[width=\textwidth]{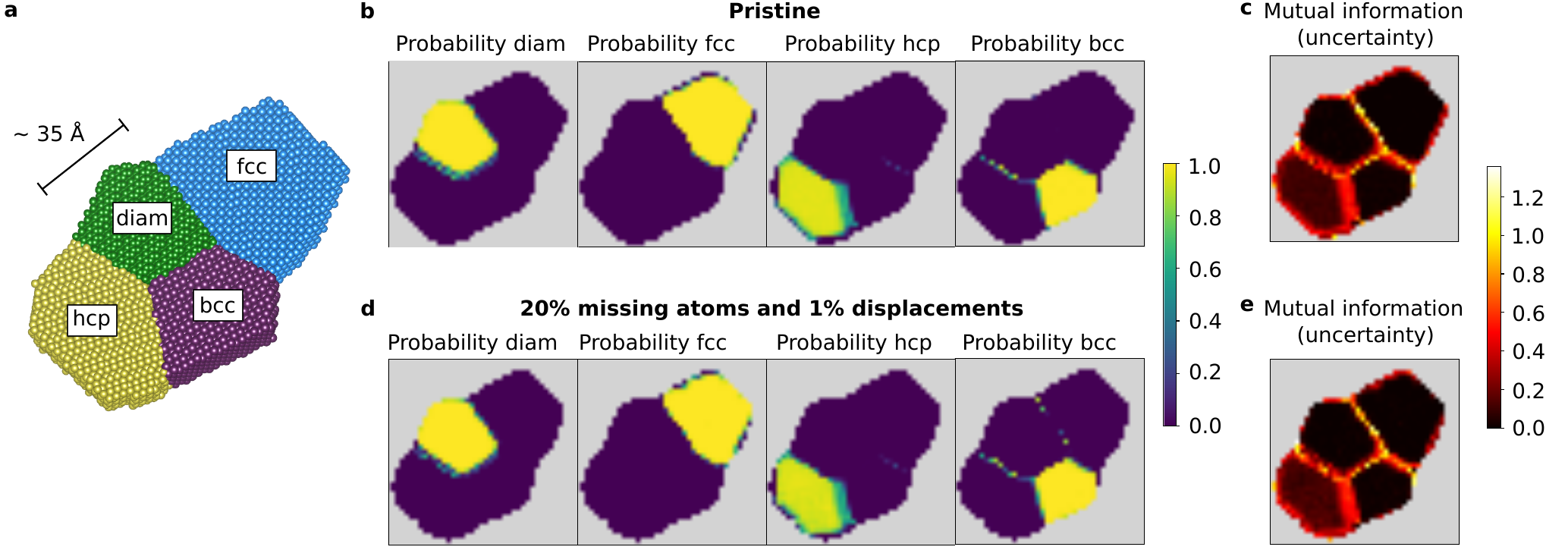}
\caption{Mono-species elemental polycrystal investigation via strided pattern matching using lower resolution (stride of 3.0 \AA in both $x$ and $y$ direction opposed to 1.0 \AA\ as in Fig. \ref{fig:synthetic_polycrystal}). 
Choosing the stride is a  trade-off between computation time and resolution. 
For instance, at the grain boundary between diamond and hcp, the transition from diamond to hexagonal diamond to hcp (cf. Supplementary Fig. \ref{fig:suppl_four_grains})
are recognized in Fig. \ref{fig:synthetic_polycrystal}b, while being obscured in the presented low resolution pictures. 
}
\label{fig:low_resolution_four_grains}
\end{figure*}

%

\begin{figure*}[!htbp]
 \includegraphics[width=\textwidth]{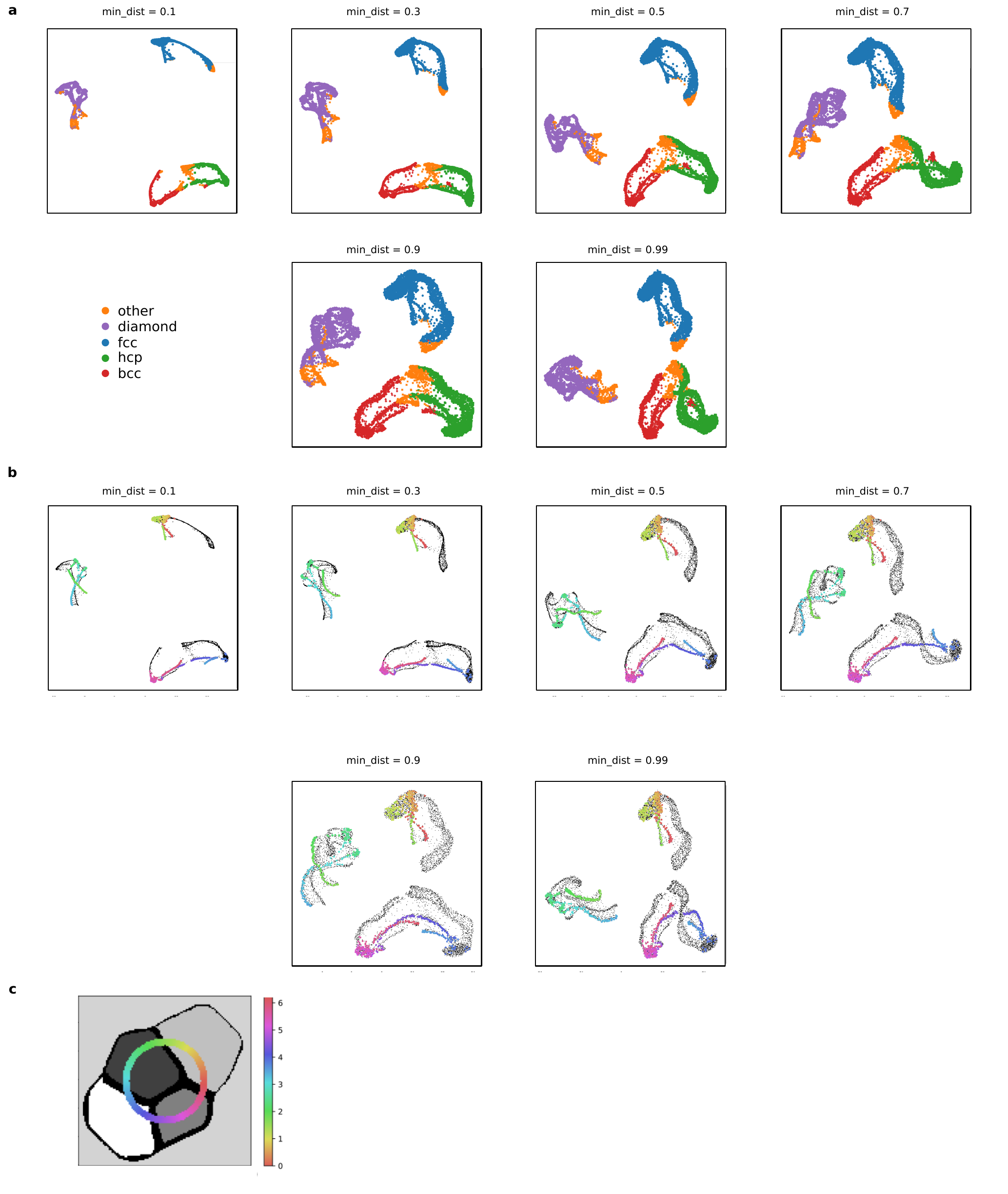}
 \caption{ Connection between UMAP embedding and real space.   
 This figure confirms the observation that ARISE's representations of different spatial regions (crystalline regions but in particular grain boundaries, here: transitions between fcc, bcc, hcp, and diamond) 
  are mapped to different regions in the UMAP projection. 
  \textbf{a}
 Influence of the $\text{min}\textunderscore\text{dist}$ parameter in the UMAP projection (number of neighbors fixed to 500). In line with intuition, for larger $\text{min}\textunderscore\text{dist}$, points appear more spread. 
 In particular, connected subregions appear in the clusters, whose connection to real space is investigated in \textbf{b}: The connected strings of points actually correspond to 
 transitions within and between crystalline regions. This is demonstrated by traversing a circle around the center of the real space structure (\textbf{c}) and coloring the embedded points according to the angle. 
 \label{fig:supp_hdbscan_pos_gb}}
\end{figure*}

\begin{figure*}[!htbp]
 \includegraphics[width=\textwidth]{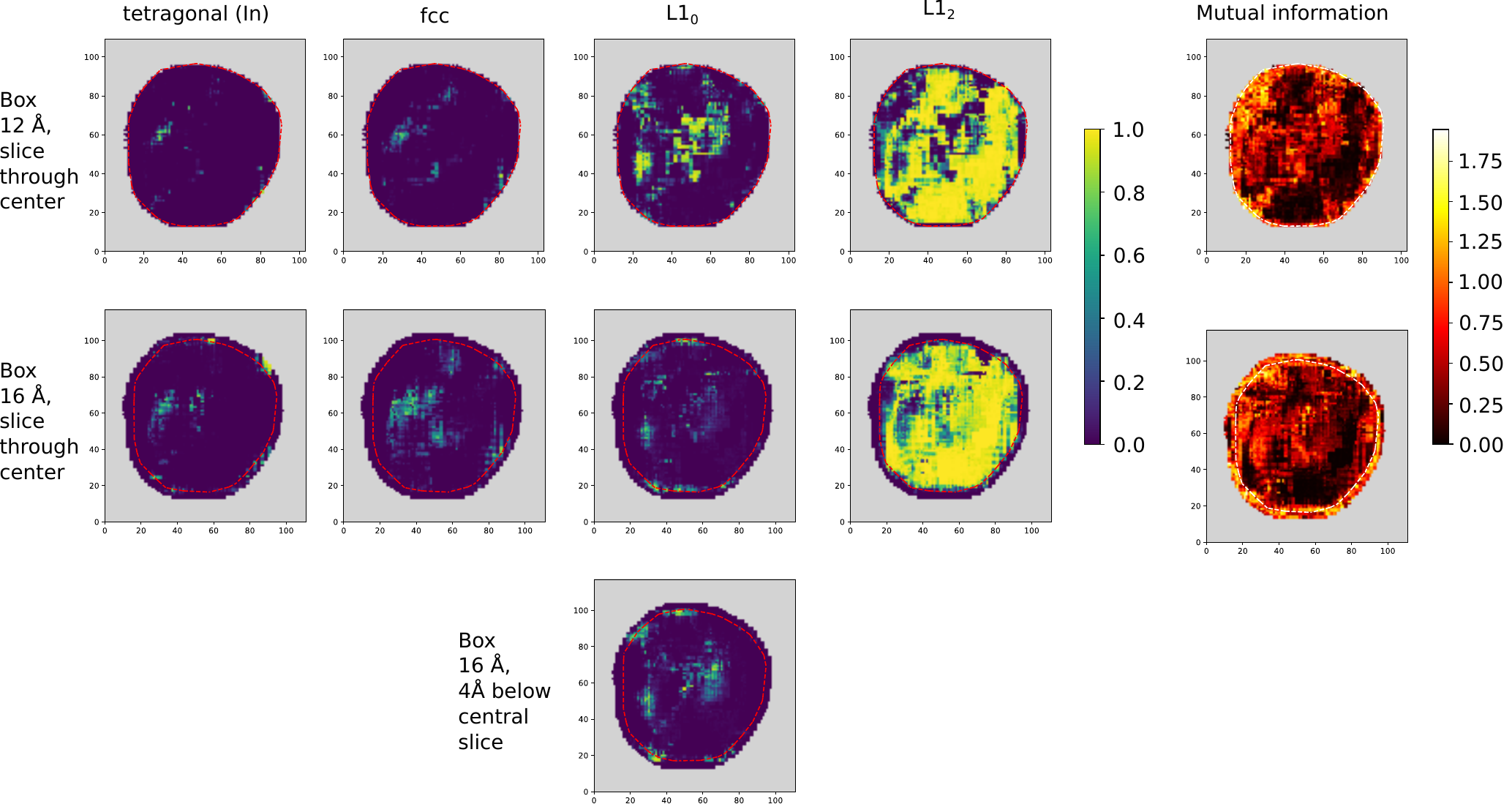}
 \caption{ 
Comparison of 
 crystal maps (slice through center, most important prototypes and mutual information) for AET nanoparticle data\cite{yang2017deciphering} for two different box sizes. 
 Dashed lines indicate the crystal boundaries in all 2D maps. 
 ARISE allows 
 to detect the appearance of the tetragonally distorted fcc prototype (In). For larger box sizes, the fcc assignment increases in the center  and also the L1$_2$ classification probability 
 rises. 
 While the central slice of the L1$_0$ prototype for a box size of 16\,\AA\ shows only weak signal, a slice slightly below
 reveals higher probability (see bottom, isolated slice), i.e., ARISE does not overlook this physically relevant phase. 
 \label{fig:supp_nanoparticle_2017_uncertainty}}
\end{figure*}

\begin{figure*}[!htb]
\centering
\includegraphics[width=\textwidth]{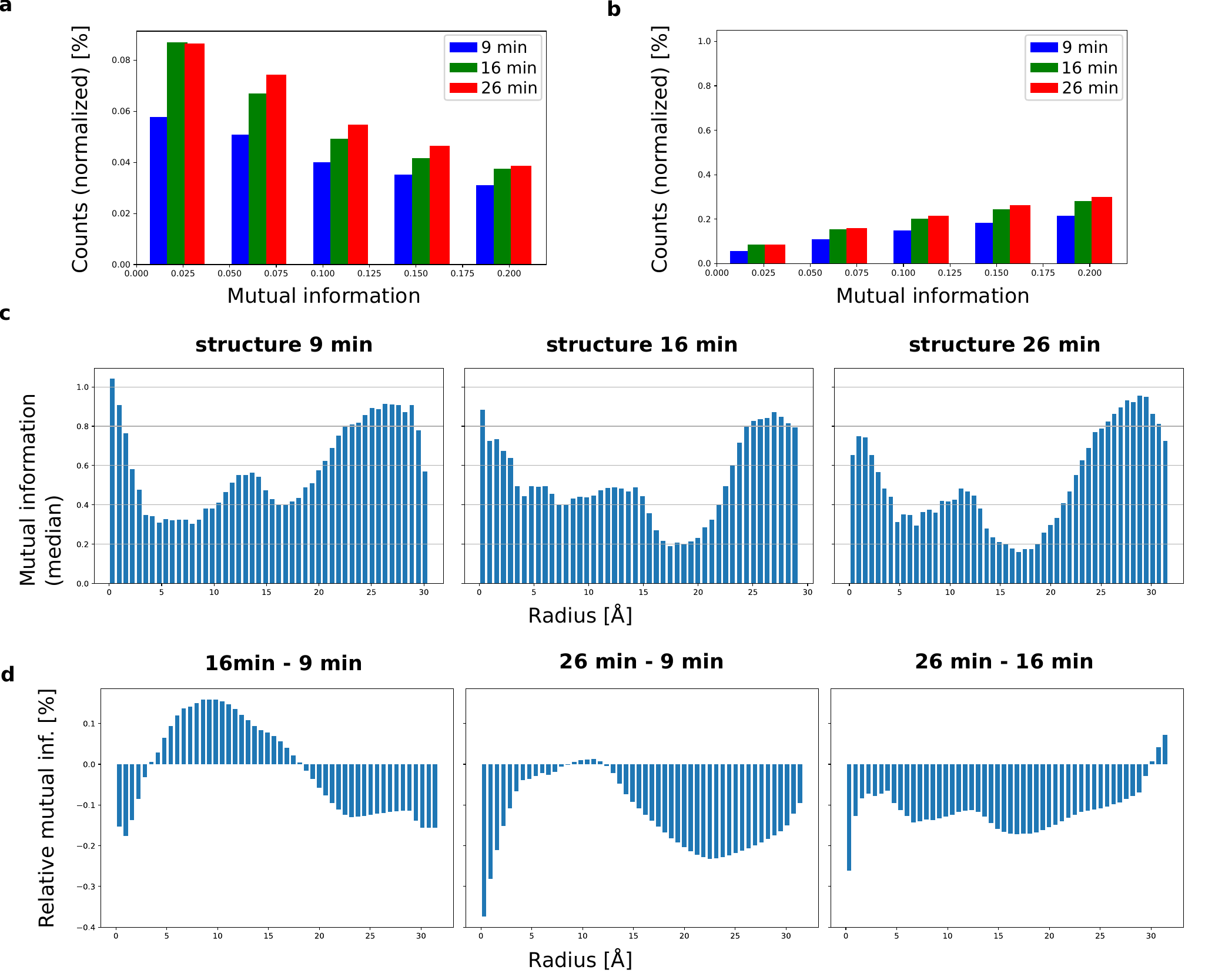}
\caption{Quantitative study of mutual information distribution for different annealing times.  
\textbf{a} Histogram of mutual information values for each annealing time (where the corresponding histograms are normalized via dividing each bin by the total number of boxes). 
Only mutual-information values smaller than 0.2 are shown, which correspond to the ``dark'', i.e., low mutual information spots in Fig.  \ref{fig:AET}c. 
\textbf{b} Cumulative distribution calculated for the histogram shown in \textbf{a}. 
From \textbf{a}, \textbf{b} it is apparent that the number of low-uncertainty boxes increases for larger annealing times. 
\textbf{c-d} Investigation of the radial distribution of the mutual information. 
\textbf{c} Histograms of uncertainty (mutual information) obtained via spatially binning the SPM maps of \ref{fig:AET}\textbf{c}  into spherical shells,  
where the median is computed for each bin. Given a mutual information value, the associated radius is calculated 
as the distance of the center of the corresponding box (as obtained via SPM) to the center of the most central box. 
\textbf{d} Each panel shows the difference between the cumulative distributions of two annealing times, where the cumulative distributions are calculated 
from  the histograms shown in \textbf{b}. In addition the histograms are normalized the following way: 
Given the times $t_1, t_2$ with $t_1<t_2$, the cumulative sum of $t_2-t_1$ is calculated and then divided by the cumulative sum of time $t_1$
such that the fractional change from $t_1$ to $t_2$ is obtained. 
One can  conclude that in \textbf{c} a clear decrease of mutual information can be spotted in specific regions, e.g., for 
the radial region 15-20\,\AA. 
The cumulative sums that are used in \textbf{d} allow to quantify the order more globally in the sense that each bin (of the cumulative sum corresponding to a specific annealing time) is proportional to the spherically averaged  integral 
from  radius zero up to the radius corresponding to the bin. 
Since the particle sizes are changing over time due to diffusion, the particles have different size. 
Thus, we single out a radius at which to compare the global order: for instance comparing the bins corresponding to a radius of r=25\,\AA, we see that for all three panels, 
the values are negative and thus the structure that has been annealed longer shows larger global order. 
}
\label{fig:AET_annealing_supp}
\end{figure*}

\clearpage
\clearpage

\begin{table}

 \begin{center}
\begin{tabular}{lllll}
 \# & Prototype & Symmetry & Material type & Data source \\
  \hline \hline
 1.&  bcc (W) & 229, cubic & Bulk, Elemental solid & AFLOW / NOMAD \\
 2.&  diamond (C) & 227, cubic & Bulk, Elemental solid & AFLOW / NOMAD \\
 3.&  fcc (Cu) & 225, cubic  & Bulk, Elemental solid & AFLOW / NOMAD \\
 4.&  $\alpha$-Po & 221, (simple) cubic & Bulk, Elemental solid & AFLOW / NOMAD \\
 5.&  hcp (Mn) & 194, hexagonal & Bulk, Elemental solid & AFLOW / NOMAD \\
 6.&  $\alpha$-La (dhcp) & 194, hexagonal & Bulk, Elemental solid & AFLOW / NOMAD \\
 7.&  Hexagonal diamond & 194, hexagonal & Bulk, Elemental solid & AFLOW / NOMAD \\
 8.&  Hexagonal graphite & 194, hexagonal & Bulk, Elemental solid & AFLOW / NOMAD \\
 9.&  Sn & 191, (simple) hexagonal & Bulk, Elemental solid & AFLOW / NOMAD \\
 10.&  Buckled graphite & 186, hexagonal & Bulk, Elemental solid & AFLOW / NOMAD \\
 11.&  $\alpha$-As & 166, rhombohedral & Bulk, Elemental solid & AFLOW / NOMAD \\
 12.&  $\alpha$-Hg & 166, rhombohedral & Bulk, Elemental solid & AFLOW / NOMAD \\
 13.& $\alpha$-Sm & 166, rhombohedral & Bulk, Elemental solid & AFLOW / NOMAD \\
 14.&  $\beta$-O & 166, rhombohedral & Bulk, Elemental solid & AFLOW / NOMAD \\
 15.&  $\beta$-Po & 166, rhombohedral & Bulk, Elemental solid & AFLOW / NOMAD \\
 16.& $\gamma$-Se & 152, trigonal hexagonal & Bulk, Elemental solid & AFLOW / NOMAD \\
 17.&  Rhombohedral graphite & 166, rhombohedral & Bulk, Elemental solid & AFLOW / NOMAD \\
 18.&  $\alpha$-Pa & 139, (body-centered) tetragonal & Bulk, Elemental solid & AFLOW / NOMAD \\
 19.&  $\beta$-Sn & 141, (body-centered) tetragonal & Bulk, Elemental solid & AFLOW / NOMAD \\
 20.&  In & 139, (body-centered) tetragonal & Bulk, Elemental solid & AFLOW / NOMAD \\
 21.&  $\gamma$-N & 136, (simple) tetragonal & Bulk, Elemental solid & AFLOW / NOMAD \\
 22.&  $\beta$-Np & 129, (simple) tetragonal & Bulk, Elemental solid & AFLOW / NOMAD \\
 23.&  $\gamma$-Pu & 70, (face-centered) orthorhombic & Bulk, Elemental solid & AFLOW / NOMAD \\
 24.&  $\alpha$-Ga & 64, (base-centered) orthorhombic & Bulk, Elemental solid & AFLOW / NOMAD \\ 
 25.&  Black phosphorus & 64, (base-centered) orthorhombic & Bulk, Elemental solid & AFLOW / NOMAD \\
 26.&  Molecular iodine & 64, (base-centered) orthorhombic & Bulk, Elemental solid & AFLOW / NOMAD \\
 27.&  $\alpha$-U & 63, (base-centered) orthorhombic & Bulk, Elemental solid & AFLOW / NOMAD \\
  \end{tabular}
 \end{center}
 \caption{Complete list of prototypes (part I) included in the training set of this work. If provided by the respective resources, information on space group, crystal system or Bravais lattice is listed. \label{table:prototype_listing_part_I}}
 \end{table}
  \clearpage
  \clearpage
  \begin{table}
  \begin{center}
 \begin{tabular}{lllll}
  \# & Prototype & Symmetry & Material type & Data source \\
  \hline \hline
   28.& NaCl & 225, cubic & Bulk, Binary compound & AFLOW / NOMAD \\
 29.& CsCl & 221, cubic & Bulk, Binary compound & AFLOW / NOMAD \\
 30.& L1$_2$ (Cu$_3$Au) & 221 (simple) cubic & Bulk, Binary compound & AFLOW / NOMAD \\
 31.& Zinc blende (ZnS) & 216, (face-centered) cubic & Bulk, Binary compound & AFLOW / NOMAD \\
 32.& FeSi & 198 (simple) cubic & Bulk, Binary compound & AFLOW / NOMAD \\
  33.& NiAs & 194, hexagonal & Bulk, Binary compound & AFLOW / NOMAD \\
  34.& Wurtzite (ZnS) & 186, hexagonal & Bulk, Binary compound & AFLOW / NOMAD \\
   35.& L1$_0$ (CuAu) & 123, (simple) tetragonal & Bulk, Binary compound & AFLOW / NOMAD \\
  36.& CrB & 63, (base-centered) orthorhombic & Bulk, Binary compound & AFLOW / NOMAD \\
  37.& MnP & 62, (simple) orthorhombic & Bulk, Binary compound & AFLOW / NOMAD \\
 38.& FeB & 62, (simple) orthorhombic & Bulk, Binary compound & AFLOW / NOMAD \\
 39.& AgNbO$_3$ & cubic & Bulk, Ternary compound & CMR \\
 40.& CsSnI$_3$ & cubic & Bulk, Ternary compound & CMR \\
 41.& CsSnCl$_3$ & tetragonal & Bulk, Ternary compound & CMR \\
 42.& Cs$_2$WO$_4$ & tetragonal & Bulk, Ternary compound & CMR \\
 43.& Ca$_3$Ge$_2$O$_7$ & tetragonal & Bulk, Ternary compound & CMR \\
  44. & CsSnCl$_3$ & orthorhombic & Bulk, Ternary compound & CMR \\
  45.& Cu$_2$BaGeSe$_4$ & 144 (trigonal) & Bulk, Quaternary compound & CMR \\
 46. & Cu$_2$CdSnS$_4$ & 121 (tetragonal) & Bulk, Quaternary compound & CMR \\
 47.& Cu$_2$ZnSnS$_4$ & 82 (tetragonal) & Bulk, Quaternary compound & CMR \\
  48.& Cu$_2$KVS$_4$ & 40 (orthorhombic) & Bulk, Quaternary compound & CMR \\
   49.& Cu$_2$CdGeS$_4$ & 31 (orthorhombic)  & Bulk, Quaternary compound & CMR \\
  50.& Cu$_2$ZnSiS$_4$ & 7 (monoclinic)  & Bulk, Quaternary compound & CMR \\
  \end{tabular}
 \end{center}
 \caption{Complete list of prototypes (part II) included in the training set of this work. \label{table:prototype_listing_part_II}}
 \end{table}
  \clearpage
  \clearpage

  \begin{table}
  \begin{center}
 \begin{tabular}{lllll}
  \# & Prototype & Symmetry & Material type & Data source \\
  \hline \hline
   51. & Graphene & 191 (hexagonal) & 2D Materials & CMR \\
52. & Ti$_3$C$_2$ & 187 (hexagonal) & 2D Materials & CMR \\
53. & Ti$_3$C$_2$O$_2$ & 187 (hexagonal) & 2D Materials & CMR \\
54. & MoS$_2$ & 187 (hexagonal) & 2D Materials & CMR \\
55. & Ti$_3$C$_2$H$_2$O$_2$ & 187 (hexagonal) & 2D Materials & CMR \\
56. & GaS & 187 (hexagonal) & 2D Materials & CMR \\
57. & BN & 187 (hexagonal) & 2D Materials & CMR \\
58. & Ti$_2$CH$_2$O$_2$ & 164 (trigonal) & 2D Materials & CMR \\
59. & Ti$_2$CO$_2$ & 164 (trigonal) & 2D Materials & CMR \\
60. & CdI$_2$ & 164 (trigonal) & 2D Materials & CMR \\
61. & CH & 164 (trigonal) & 2D Materials & CMR \\
62. & CH$_2$Si & 156 (trigonal) & 2D Materials & CMR \\
63. & Ti$_4$C$_3$ & 156 (trigonal) & 2D Materials & CMR \\
64. & BiTeI & 156 (trigonal) & 2D Materials & CMR \\
65. & Ti$_4$C$_3$O$_2$ & 156 (trigonal) & 2D Materials & CMR \\
66. & GeSe & 156 (trigonal) & 2D Materials & CMR \\
67. & MoSSe & 156 (trigonal) & 2D Materials & CMR \\
68. & Ti$_4$C$_3$H$_2$O$_2$ & 156 (trigonal) & 2D Materials & CMR \\
69. & AgBr$_3$ & 150 (trigonal) & 2D Materials & CMR \\  
    \end{tabular}
 \end{center}
 \caption{Complete list of prototypes (part III) included in the training set of this work. \label{table:prototype_listing_part_II}}
 \end{table}

  \begin{table}
  \begin{center}
 \begin{tabular}{lllll}
  \# & Prototype & Symmetry & Material type & Data source \\
  \hline \hline
  70. & TiCl$_3$ & 150 (trigonal) & 2D Materials & CMR \\
71. & BiI$_3$ & 147 (trigonal) & 2D Materials & CMR \\
72. & FeSe & 129 (tetragonal) & 2D Materials & CMR \\
73. & PbSe & 123 (tetragonal) & 2D Materials & CMR \\
74. & GeS$_2$ & 115 (tetragonal) & 2D Materials & CMR \\
75. & C$_3$N & 65 (orthorhombic) & 2D Materials & CMR \\
76. & FeOCl & 59 (orthorhombic) & 2D Materials & CMR \\
77. & P & 28 (orthorhombic) & 2D Materials & CMR \\
78. & PdS$_2$ & 14 (monoclinic) & 2D Materials & CMR \\
79. & MnS$_2$ & 14 (monoclinic) & 2D Materials & CMR \\
80. & GaSe & 12 (monoclinic) & 2D Materials & CMR \\
81. & TiS$_3$ & 11 (monoclinic) & 2D Materials & CMR \\
82. & WTe$_2$ & 11 (monoclinic) & 2D Materials & CMR \\
83. & HfBrS & 7 (monoclinic) & 2D Materials & CMR \\
84. & RhO & 6 (monoclinic) & 2D Materials & CMR \\
85. & SnS & 6 (monoclinic) & 2D Materials & CMR \\
86. & NiSe & 6 (monoclinic) & 2D Materials & CMR \\
87. & AuSe & 6 (monoclinic) & 2D Materials & CMR \\  
    \end{tabular}
 \end{center}
 \caption{Complete list of prototypes (part IV) included in the training set of this work. \label{table:prototype_listing_part_II}}
 \end{table}
 
    \clearpage
  \clearpage

  \begin{table}
  \begin{center}
 \begin{tabular}{lllll}
  \# & Prototype & Symmetry & Material type & Data source \\
  \hline \hline
88. & VTe$_3$ & 6 (monoclinic) & 2D Materials & CMR \\
89. & ReS$_2$ & 2 (monoclinic) & 2D Materials & CMR \\
90. & ScPSe$_3$ & 1 (triclinic) & 2D Materials & CMR \\
91. & PbA$_2$I$_4$ & 1 (triclinic) & 2D Materials & CMR \\
92. & PbS & 1 (triclinic) & 2D Materials & CMR \\
93. & CrW$_3$S$_8$ & 1 (triclinic) & 2D Materials & CMR \\
94. & VPSe$_3$ & 1 (triclinic) & 2D Materials & CMR \\
95. & CrWS$_4$ & 1 (triclinic) & 2D Materials & CMR \\
96. & MnPSe$_3$ & 1 (triclinic) & 2D Materials & CMR \\
97. & Carbon nanotube & armchair, (3,3), $30.0^\circ, 4.07\,$\AA & Nanotubes, mono-species  & ASE \\
98. & Carbon nanotube & armchair, (4,4), $30.0^\circ, 5.42\,$\AA & Nanotubes, mono-species  & ASE \\
99. & Carbon nanotube & chiral, (4,2), $19.11^\circ, 4.14\,$\AA & Nanotubes, mono-species  & ASE \\
100. & Carbon nanotube & chiral, (4,3), $25.28^\circ, 4.76\,$\AA & Nanotubes, mono-species  & ASE \\
101. & Carbon nanotube & chiral, (5,1), $8.95^\circ, 4.36\,$\AA & Nanotubes, mono-species  & ASE\\
102. & Carbon nanotube & chiral, (5,2), $16.1^\circ, 4.89\,$\AA & Nanotubes, mono-species  & ASE\\
103. & Carbon nanotube & chiral, (5,3), $21.79^\circ, 5.48\,$\AA & Nanotubes, mono-species  & ASE\\
104. & Carbon nanotube & chiral, (6,1), $7.59^\circ, 5.13\,$\AA & Nanotubes, mono-species  & ASE\\
105. & Carbon nanotube & chiral, (6,2), $13.9^\circ, 5.65\,$\AA & Nanotubes, mono-species  & ASE\\
106. & Carbon nanotube & chiral, (7,1), $6.59^\circ, 5.91\,$\AA & Nanotubes, mono-species  & ASE\\
107. & Carbon nanotube & zigzag, (6,0), $0.0^\circ, 4.7\,$\AA & Nanotubes, mono-species  & ASE\\
108. & Carbon nanotube & zigzag, (7,0), $0.0^\circ, 5.48\,$\AA & Nanotubes, mono-species  & ASE
\end{tabular}
\end{center}
\caption{Complete list of prototypes (part V) included in the training set of this work. For the carbon nanotubes, the symmetry column specifies the configuration type (chiral, zigzag or armchair) 
together with the corresponding chiral numbers (n,m), the chiral angle $\theta$ and the nanotube diameter. 
\label{table:prototype_listing_part_III}}
\end{table}

\end{document}